\pdfoutput=1

\documentclass[journal,transmag]{IEEEtran}
\ifCLASSINFOpdf
\else
\fi

\usepackage[pdftex]{graphicx}
\usepackage{framed}
\usepackage{amsmath}
\usepackage{threeparttable}
\usepackage{booktabs}
\usepackage{subfigure}

\usepackage{color}

\begin{document}
%
\title{
Ten Years of Research on Intelligent Educational Games for Learning Spelling and Mathematics
}


\author{\IEEEauthorblockN{
}}

\author{
\IEEEauthorblockN{Barbara Solenthaler\IEEEauthorrefmark{1},
Severin Klingler\IEEEauthorrefmark{1},
Tanja K\"aser\IEEEauthorrefmark{2}, and
Markus Gross\IEEEauthorrefmark{1}}
\IEEEauthorblockA{\IEEEauthorrefmark{1}ETH Zurich, Departement of Computer Science, Switzerland}
\IEEEauthorblockA{\IEEEauthorrefmark{2}Stanford University, AAA Lab, Graduate School of Education, USA}
}

%



\IEEEtitleabstractindextext{%
\begin{abstract}
In this article, we present our findings from ten years of research on intelligent educational games. 
We discuss the architecture of our training environments for learning spelling and mathematics, and specifically focus on the representation of the content and the controller that enables personalized trainings.
We first show the multi-modal representation that reroutes information through multiple perceptual cues and discuss the game structure. We then present the data-driven student model that is used for a personalized, adaptive presentation of the content. We further leverage machine learning for analytics and visualization tools targeted at teachers and experts. 
A large data set consisting of training sessions of more than 20,000 children allows statistical interpretations and insights into the nature of learning.
\end{abstract}

}

\maketitle

\IEEEdisplaynontitleabstractindextext

%
\IEEEpeerreviewmaketitle


In Intelligent Tutoring Systems (ITS), computer-based learning and educational games harness the motivational power of digital content and games in order to engage students to practice and learn a specific topic. These educational platforms have to be carefully designed, considering educational principles as well as optimal task sequences in order to foster maximal learning progress. 
In recent years, awareness was raised in the ITS community that the use of gamified elements is not sufficient to foster maximal learning progress. According to Csikszentmihalyi's Flow theory~\cite{Csi90}, students experience boredom if tasks are repetitive or if they do not adhere to the student's current knowledge level. 
In order to cope with that problem, the traditional architecture
was extended by a machine learning module that allows to capture student properties and to react accordingly. In its simplest form, the module captures the student's performance and adapts the task difficulty correspondingly.
Student behavior is, however, very complex and goes beyond cognition. Therefore, more sophisticated models used in ITS - such as the one we are presenting in this article - include machine learning algorithms to capture the fine-grained and entangled dynamics of student behavior. This enables a personalized, adaptive presentation of the content, making learning more effective and also more fun for a student.

Adaptivity of the content is particularly important in game-based therapy systems for children with learning disabilities. The digital environment presents an inexpensive extension to conventional one-to-one therapy and provides a fear-free learning environment to a child who has also often developed an aversion to the subject. The AI controller is the key in such systems since it reacts on the individual needs of the child and hence optimizes the child's overall learning progress and attitude towards the learning. 

In this article, we present an architecture for intelligent educational games targeted at children with such learning difficulties, in particular we focus on difficulties in spelling and mathematics (developmental dyslexia and dyscalculia). The architecture has been developed and refined over ten years of research, and we discuss the core methods and findings in the following.
The central concept of our training environments is a multi-modal representation using a set of codes, rerouting information about letters and numbers through multiple perceptual cues (such as topology, color, shape, and auditory representations). The two training environments feature 3D graphics and interaction components and thus allow immersion in a playful 3D world. The different information channels interact with the user and give auditory and visual feedback if errors occur. The displayed games, tasks and difficulty levels are individualized by using machine learning strategies to virtually represent a student and capturing complex user behavior such as cognition, affect, and student traits. Machine learning is further used to provide visual analytics tools for teachers and experts.
Our large data set consisting of training data of more than 20,000 children allows us to get insights into the nature of learning.

\section{Architecture Overview}
\begin{figure}[t!]%
\centering 

\includegraphics[trim=0.5cm 8cm 0.5cm 8cm, clip=true, width=1.0\columnwidth]{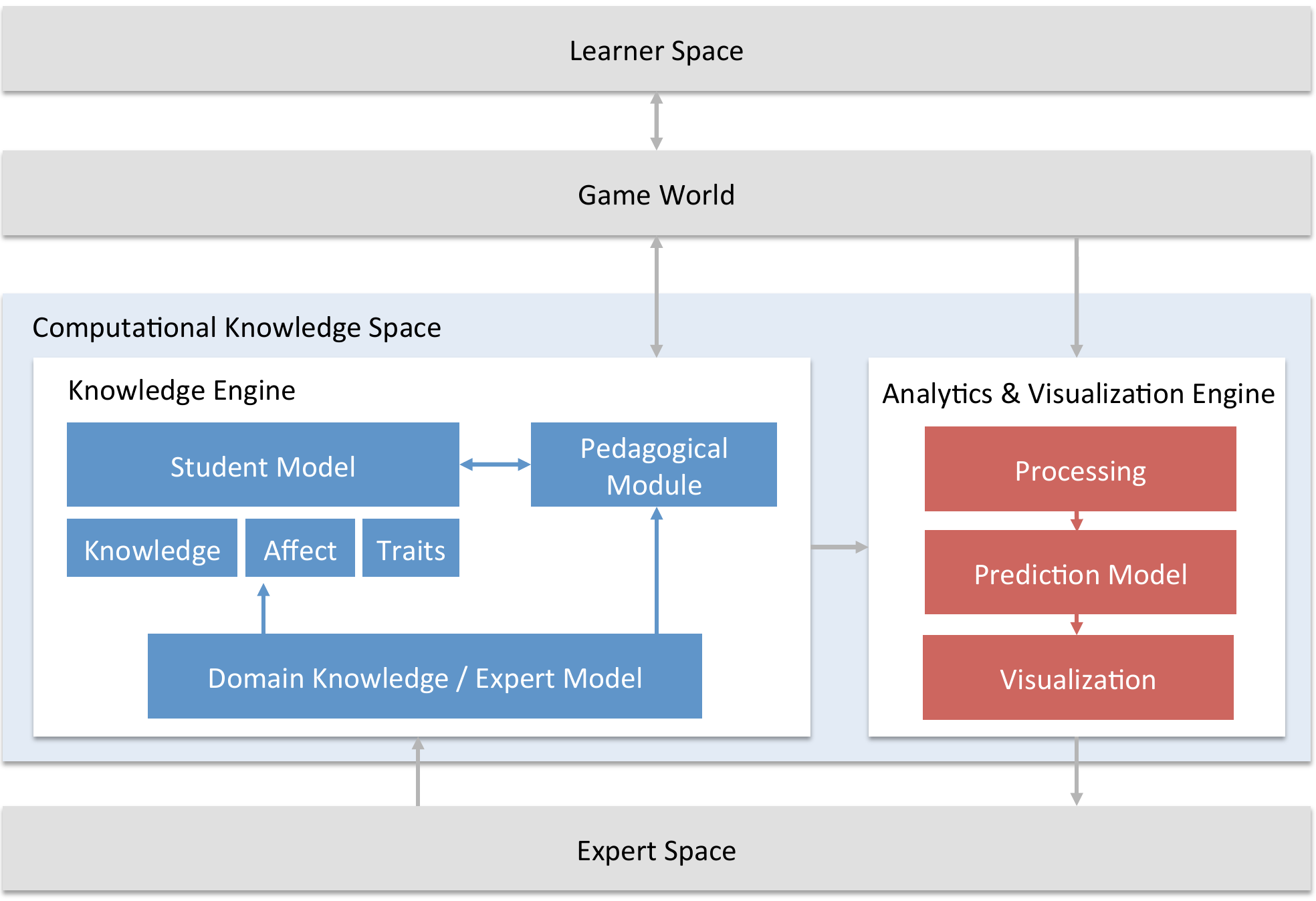}
\caption{
Our architecture for the two educational games for learning spelling and mathematics. The Computational Knowledge Space represents the machine learning module for student modeling, personalized game adaptation and expert reasoning.
}
\label{fig:architecture}
\end{figure}
Our architecture is illustrated in Figure~\ref{fig:architecture} and consists of four components: The learner space (the student / user), the game world (games and tasks), the machine learning module - here referred to as the \emph{Computational Knowledge Space} - and the expert space (domain expert, psychologist; could be the game designer in the context of video games).
In this article, we focus on the \emph{Computational Knowledge Space} which is responsible for virtually modeling a student and adapting the presented content (i.e., game tasks and difficulty levels), as well as for providing an analytics tool for experts to reason about the data and to improve the training environment. We therefore divide the module into the \emph{Knowledge Engine} and the \emph{Analytics and Visualization Engine}.

The \emph{Student Model} is part of the \emph{Knowledge Engine} and can be seen as a mathematical representation of the student. It leverages machine learning techniques for predicting short- and long-term knowledge and performance, the affective state of a student and various student traits. 
The \emph{Pedagogical Module} is the controller that makes the teaching decisions and adapts the games and instructions based on the actual learning capabilities and preferences of the student.
The \emph{Domain Knowledge / Expert Model} is included in this process to optimally represent skills and learning behavior. 
The \emph{Analytics and Visualization Engine} uses again machine learning and represents the data visually. This allows experts, teachers and researchers to analyze and interpret the data, to evaluate the models, and to improve the \emph{Knowledge Engine} accordingly. 

All modules included in our architecture are commonly used in ITS and are subject to ongoing research. 
We have studied the different components for our two training environments \emph{Orthograph}~\cite{KM07,Bas10b,baschera2010poisson} and \emph{Calcularis}~\cite{KA12,KA13b,KA13d}, which we have developed for children with difficulties in learning spelling and mathematics, respectively. In the remainder of this article, we discuss each module of the \emph{Computational Knowledge Space} and show their effectiveness in the context of our applications. 
One major design decision for our system was to log each and every keystroke of a student, which resulted in a large and powerful data set. This allowed us to develop methods for feature extraction and feature processing that can be reused by the individual modules and therefore simplifies the interplay between them. 
Currently, we have data from more than $20,000$ children who worked with the systems at Swiss schools and at home, as well as data from controlled user studies conducted in German speaking countries.

\section{Games, Domain Knowledge and Expert Model}
\label{sec:domain}
%
\begin{figure}[b!]
\centering
\includegraphics[trim=0.75cm 12cm  0.5cm 12.5cm, clip=true, width=0.95\columnwidth]{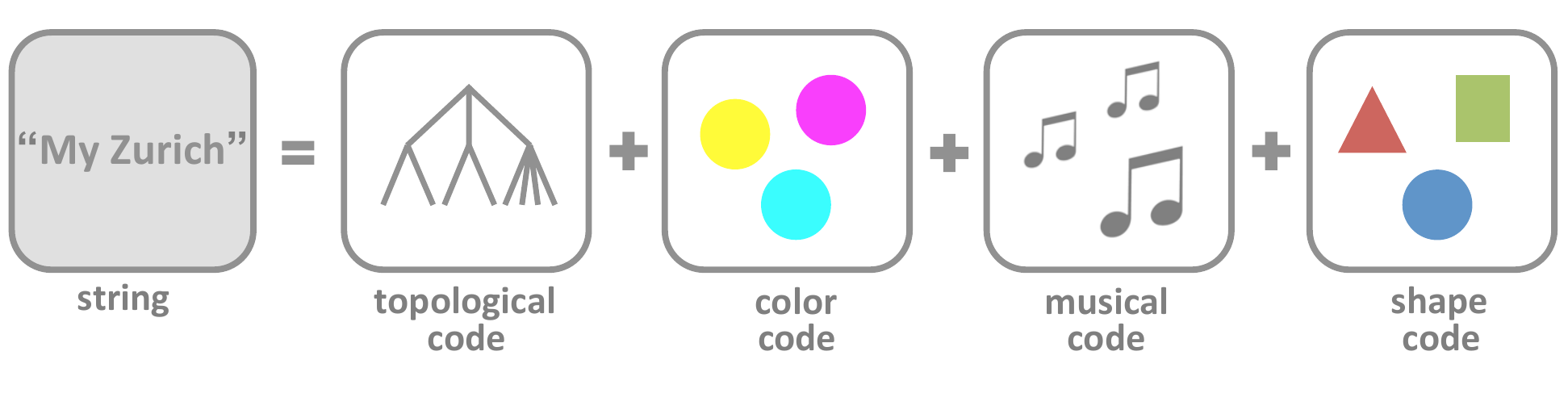}
\includegraphics[trim=3cm 5.5cm 4cm 2cm, clip=true, width=0.95\columnwidth]{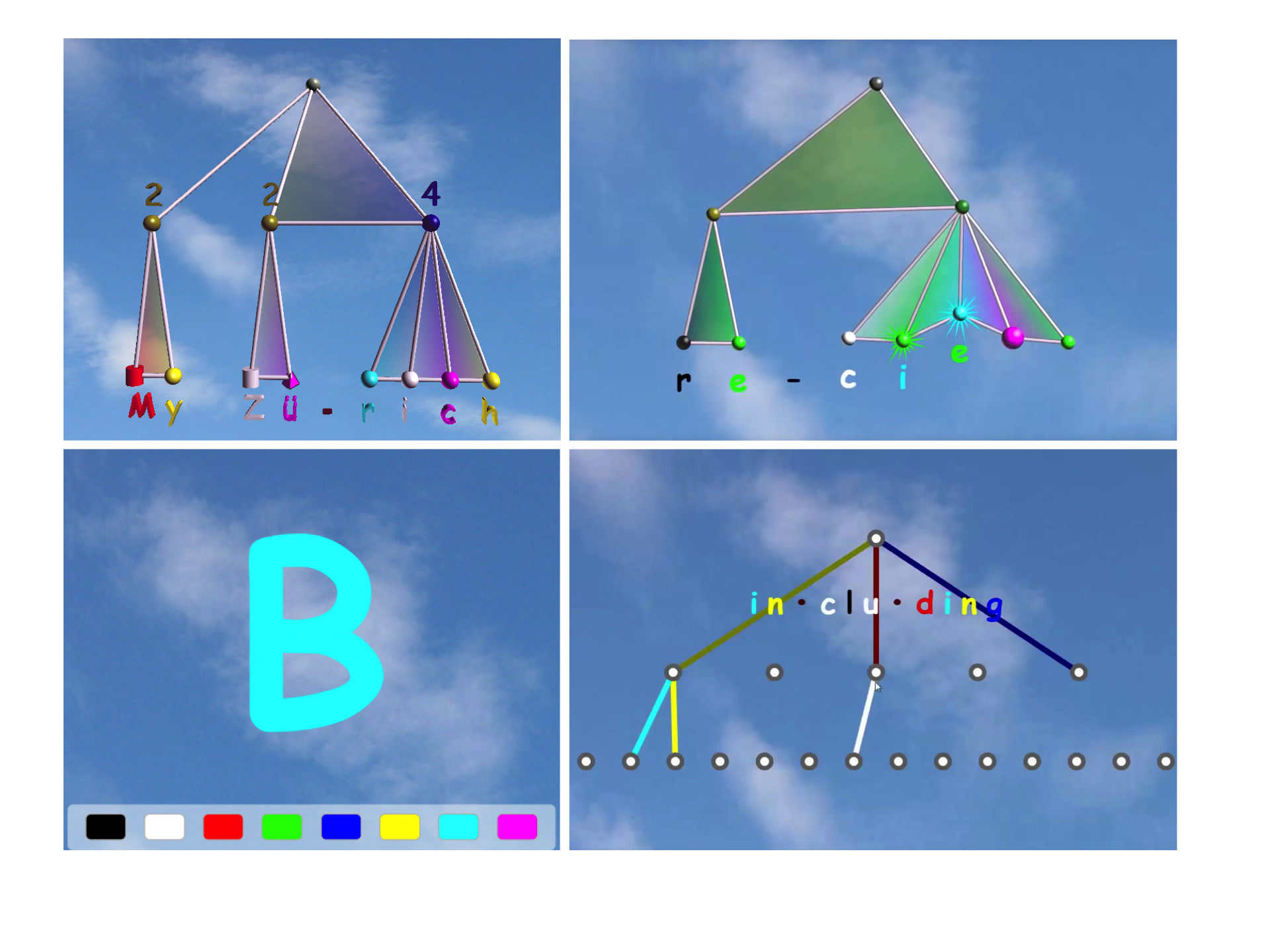}
\caption{
We encode words using multi-modal cues such as topology, color, music and shape. In Orthograph, these codes are used in the word game (top), the color game (bottom left) and the graph game (bottom right). 
}
\label{fig:gamesOrthograph}
\end{figure}
%


The two training systems for spelling learning and mathematics, \emph{Orthograph} and \emph{Calcularis}, consist of a set of instructional games for elementary school children. The games are designed based on expert knowledge and combine state-of-the-art knowledge from neuroscience and computer science, leading to a multimodal therapy for dyslexia and dyscalculia.
Multimodal information processing relates to the fact that the human brain is able to process information simultaneously through different perceptual cues and channels. The core idea of the learning is to reroute the information that is embedded in a word or number through different perceptual pathways. This goal is accomplished by encoding the information using different perceptual cues such as topology, color, music, or shape. 

The multimodal representation is illustrated for \emph{Orthograph} in Figure~\ref{fig:gamesOrthograph}, top. The graph structure shows the decomposition of a word into syllables and graphemes. The letters are represented by eight different colors; this mapping is the result of a multi-objective optimization. The idea is to associate colors with letters to eliminate mistakes due to letter confusion. The shapes are spheres for small letters, cylinders for capital letters, and pyramids for the umlauts. The auditory code computes a word-specific melody, which is played when the word is entered. The encoding is applied to the color game, the graph game, and the word game depicted in Figure~\ref{fig:gamesOrthograph}.

\begin{figure}[t!]
\centering
\includegraphics[trim=0.5cm 12.5cm 0cm 12.5cm, clip=true, width=0.95\columnwidth]{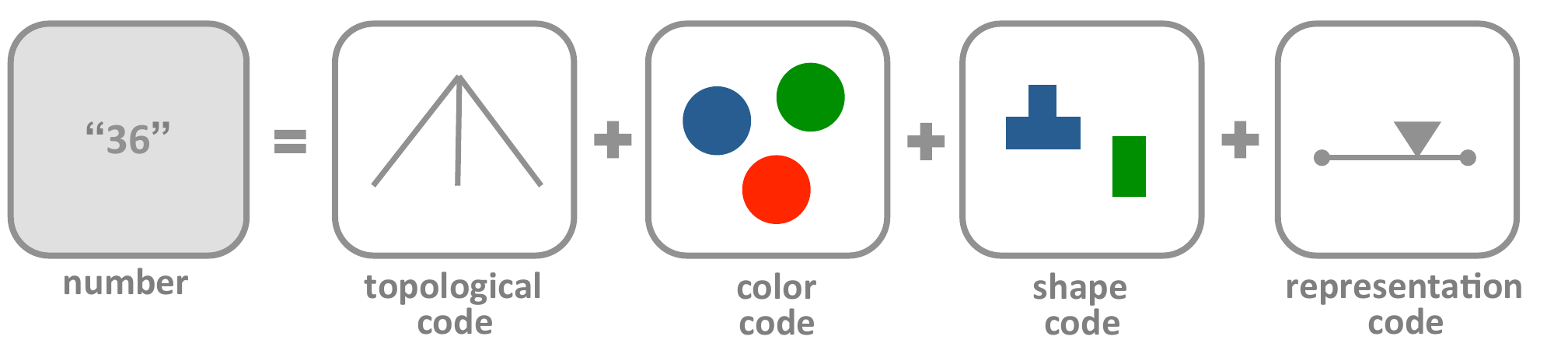}
\includegraphics[trim=0.5cm 9.5cm 0cm 9.5cm, clip=true, width=0.95\columnwidth]{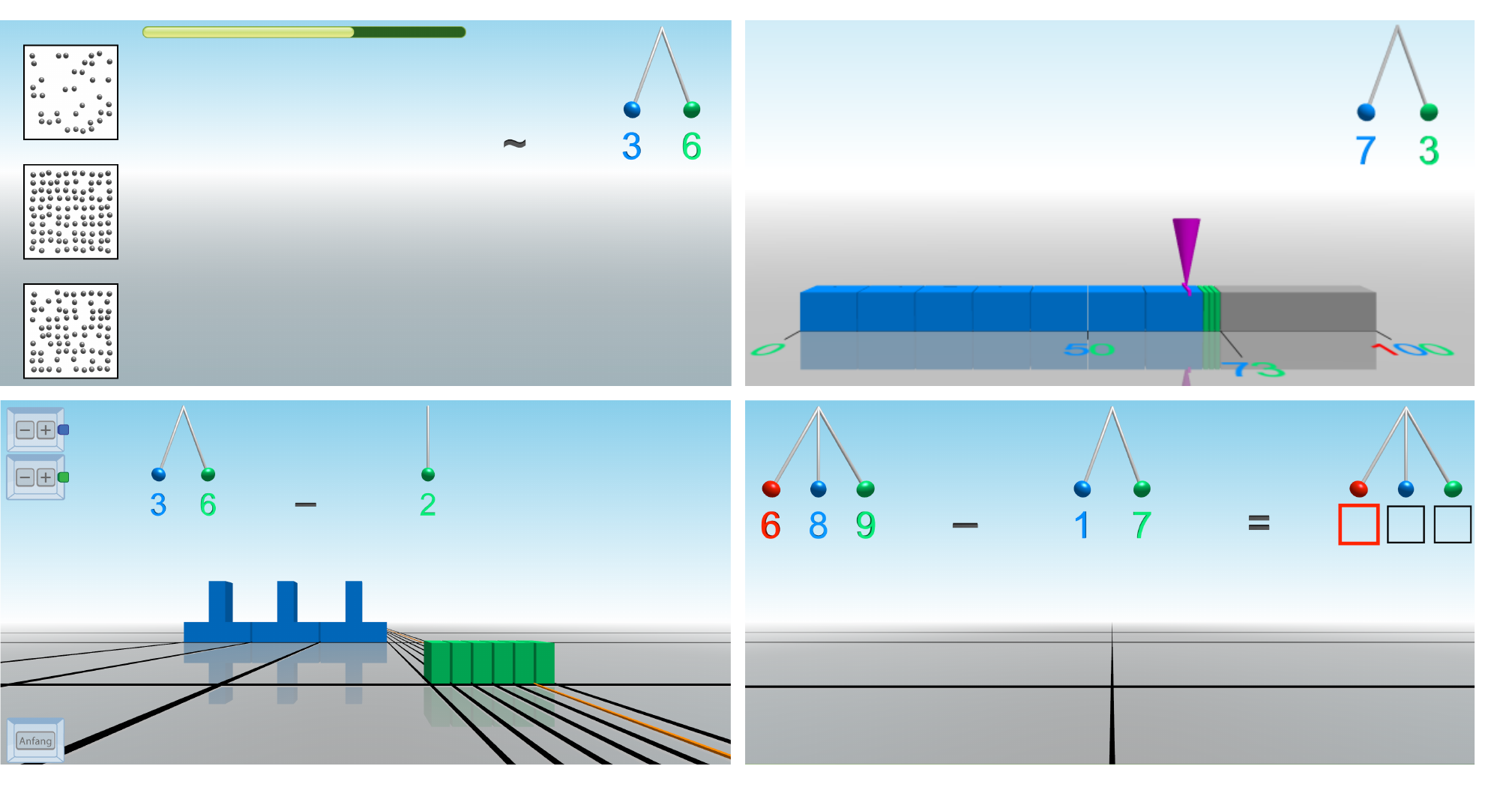}
\caption{
A multimodal encoding is used for the games in Calcularis, such as the estimation game (top left), landing game (top right), plus-minus game (bottom left) and calculator game (bottom right).
}
\label{fig:gamesCalcularis}
\end{figure}

\emph{Calcularis} uses the same idea and represents the properties of numbers using a set of codes as illustrated in Figure~\ref{fig:gamesCalcularis}, top. The place-value system is enhanced by using differently colored blocks denoting the positions of the digits and by attaching the digits to the branches of a number graph. In addition, numbers are represented as a composition of blocks with different colors indicating hundred, tens or individual units. The number line representation with integrated blocks enhances the ordinality of numbers. The codes are used in the various games, such as the estimation game, the landing game, the plus-minus game and the calculator game shown in Figure~\ref{fig:gamesCalcularis}.
The multi-modal encoding aims at enhancing the different properties of the domain-specific stimuli and facilitating word and number understanding. Moreover, the transfer of information through different channels stimulates perception and facilitates the retrieval of memory~\cite{Sha08}.

%
%

We use a sequential representation of the skill set in \emph{Orthograph} by ordering the words into groups based on increasing word difficulty. The individual skills are then trained using multiple games and tasks. 
Skills in learning mathematics depend hierarchically on the previously acquired knowledge. Our game structure follows the four-step developmental model~\cite{von2007number}. Starting from a (probably) inherited core-system representation of cardinal magnitude (step 1), the linguistic symbolization (spoken number) develops during pre-school time (step 2). The Arabic symbolization (written number) is then taught in school (step 3) and finally the analogue magnitude representation (number as a position on a number line) develops (step 4). \emph{Calcularis} further includes games that train arithmetic operations at different difficulty levels, whereas difficulty is defined by the magnitude of the numbers included in the task, the complexity of the task as well as the means allowed to solve the task.    
\section{Student Model}

The student model represents and predicts the knowledge of a student. Extended models may also include the representation and prediction of affective states, subgroup information, student characteristics, and long-term performance. The predictions of the student model are used by the pedagogical module to make teaching decisions, i.e. to adapt the presented content to the needs of the specific user.   

\subsection{Knowledge}
\label{sec:knowledge}
%
\begin{figure}[b!]
\centering
\subfigure[Hierarchical skill net of \emph{Calcularis} (Part A and B are connected as well).]
{\includegraphics[trim=0cm 0cm 0cm 0cm, clip=true, width=1\columnwidth]{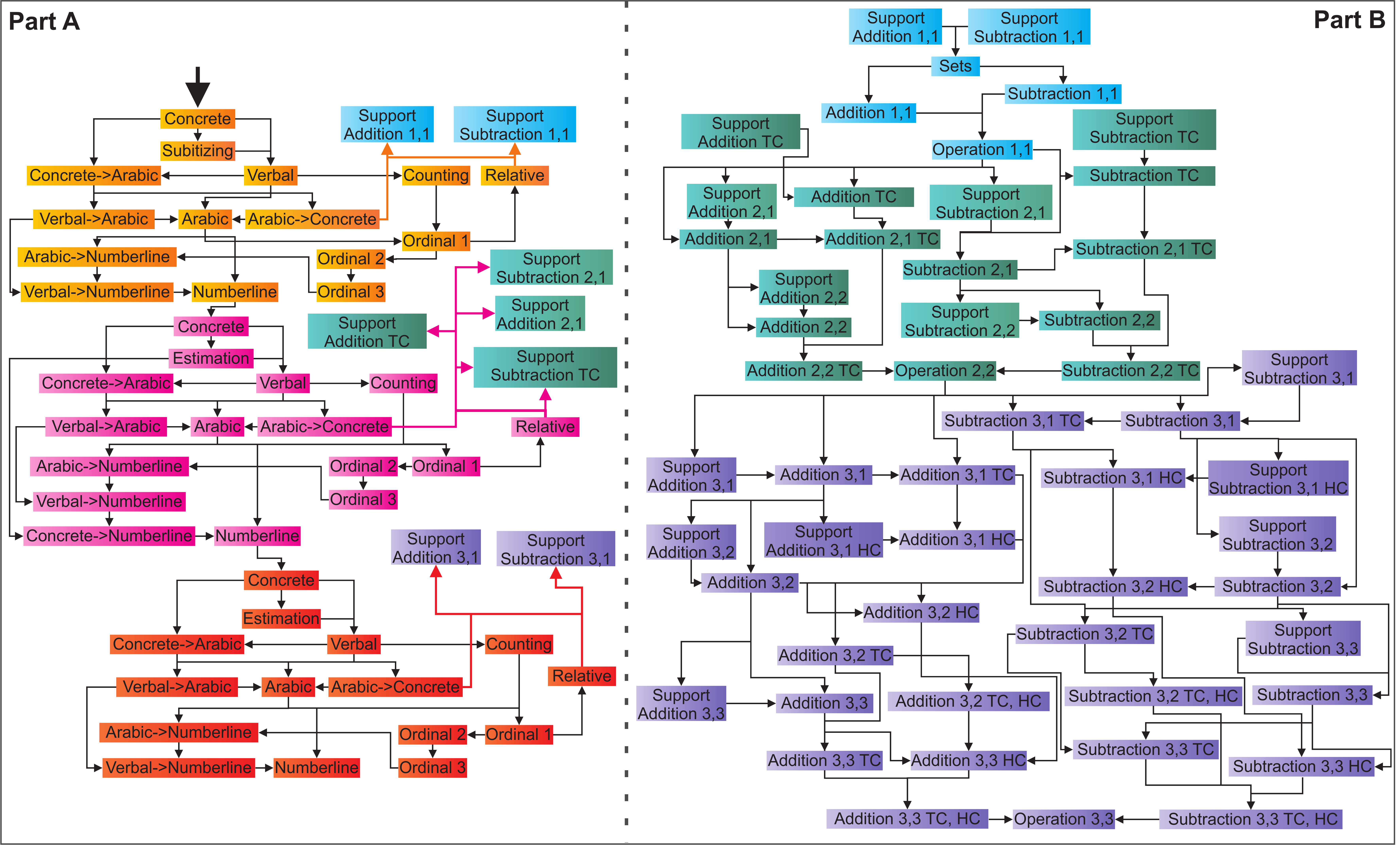}
\label{fig:skillnet}
}
\subfigure[Individualized learning paths through the skill net.]
{\includegraphics[trim=0.8cm 0.8cm 0.8cm 0.8cm, clip=true, width=0.85\columnwidth]{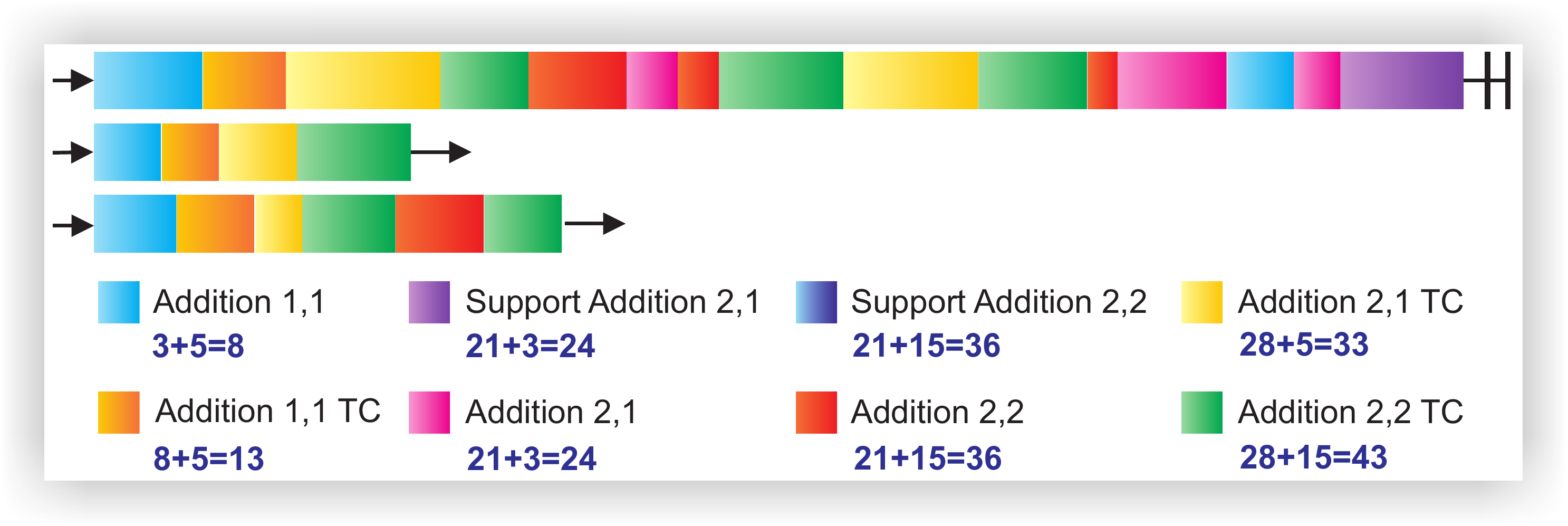}
\label{fig:learningpaths}
}
\caption{(a) The DBN used in \emph{Calcularis} is represented by a skill net that considers current neuro-cognitive findings. 
The colors denote the different number ranges 0-10 (yellow, blue), 0-100 (pink, green), and 0-1000 (red, purple). (b) The student model allows individualized training paths through the skill net as shown for three students. 
The color indicates the arithmetic task (TC= ten crossing) and the length of the rectangle the number of trials.
}
\end{figure}
Both our systems work with probability distributions: given many observations, what are the probabilities of the unknown variables. Examples for unknown variables are the student knowledge, i.e., whether a student has learned a skill $s$ at time $t$, and the affective state of the user, i.e., whether a student is attentive at time $t$.   

For adapting the training in learning spelling, we use a taxonomy of errors that a student can make and corresponding features to detect them. 
%
We first identify patterns and similarities in spelling errors across the entire word database and represent them in as few error production rules as possible, which we refer to as \emph{mal-rules}~\cite{baschera2010poisson} (capitalization and typing errors, letter confusion, phoneme-grapheme matching, phoneme omission, insertion and transposition). From the student input we can extract how much each mal-rule is activated in an error. The error behavior of a student is then described by the occurrence of the error and how often the error was committed by the student. This information serves as the input to the inference algorithm that estimates the student's difficulties with each individual mal-rule. The estimates of the individual mal-rules are continuously updated after each entered word during the training. These values are used to compute the probability that a certain mal-rule is the source of the committed error. We further compute the error expectation estimates for each word. The word selection controller in the pedagogical module then uses this information to adapt the training (Section~\ref{sec:pedagogical}). 
The conducted user studies have demonstrated a substantial improvement in spelling of 30\% after twelve weeks of training~\cite{Gro07}.

\emph{Calcularis} models the mathematical knowledge of the student using a dynamic Bayesian network (DBN).
This network consists of a directed acyclic graph
representing different mathematical skills and the relationships between them. The student model of \emph{Calcularis} consists of $100$ different skills and is illustrated in Figure~\ref{fig:skillnet}. All games of the system are associated with one (or several) skills of this network. 
Each skill can have two states: a learned state and an unlearned state. The probability that a skill is in the learned state is inferred by observing student answers to tasks associated with this skill. At the beginning of the training, all probabilities are initialized to $0.5$ as we have no knowledge about the proficiency of a learner - this is in accordance with the principle of maximum entropy. During training, the system updates the probabilities for the different skills after each solved task of the student.

Based on these skill probabilities, the controller in the pedagogical module selects the next task for the student to solve (Section~\ref{sec:pedagogical}).
%
In conducted user studies we have identified a learning progress in mathematics of about 23\%~\cite{KA12}.

Exact inference in DBNs is generally not computationally tractable, however, recently~\cite{Sch12} showed that a convex approximation allows for efficient parameter learning in a DBN. We have further shown in~\cite{kae14a} that a constrained latent structured prediction approach for parameter learning yields accurate and interpretable results: we include a-priori domain expert knowledge via regularization with constraints into our parameter learning algorithm. Furthermore, we have demonstrated that our hierarchical knowledge representation outperforms traditional approaches to student modeling in a variety of different learning domains~\cite{kae14b}.

\subsection{Affect}

Identifying the current engagement state of the user such as boredom or lack of concentration allows for a fine-grained adaptation of the training to the specific needs of the user. While previous work used various sensors (such as camera, eye-tracking, EEG monitoring headsets, bio-sensors), we investigated if the affective state of a student can be identified based on input data only. We have developed a method for modeling engagement dynamics in spelling learning~\cite{BA11} and provided a theoretical generalization of the model to the case of mathematics learning~\cite{KA13}. 
The set of features relevant for engagement dynamics are related to input and error behavior, timing features and controller induced features. Features such as the input rate and its variance as well as timing features can indicate a lack of concentration, typically leading to many help calls and more errors. The controller induced features are important as they for example indicate the time between repetitions and thus have a direct influence on forgetting.

Feature processing is based on the assumption that emotional and motivational states come in spurts and that they affect the observed features on a short-to-medium time scale. We therefore perform a time scale separation by distinguishing between sustainable progress in the observed input behavior and other local effects, such as the influence of engagement states, and combine the two effects linearly.
We then use LASSO logistic regression with 10-fold cross-validation to estimate the relation between processed features and error repetition and hence perform a feature selection. The parameters of the logistic regression indicate how features are related to the error repetition probability (ERP). The ERP is the the probability  of committing errors associated with missing knowledge in spelling repeatedly for the same word.
In the selected features, we identified three main effects influencing the knowledge state at the next repetition:
\begin{enumerate}
\item \emph{Focused state:} Indicates if a student is focused or distracted. In a non-focused state more minor errors due to lapse of concentration occur, which are less likely to be committed again at the next repetition (lower ERP).
\item \emph{Receptive state:} Indicates the receptiveness of the student (beyond attention span). Non-receptive state inhibits learning and causes a higher ERP.
\item \emph{Forgetting:} The time (decay) and number of inputs (interference) between error and repetition induce forgetting of learned spelling and increase the ERP.
\end{enumerate}
\begin{figure}[t!]
\centering
\includegraphics[trim=0cm 0cm 0cm 1cm, clip=true, width=1.0\columnwidth]{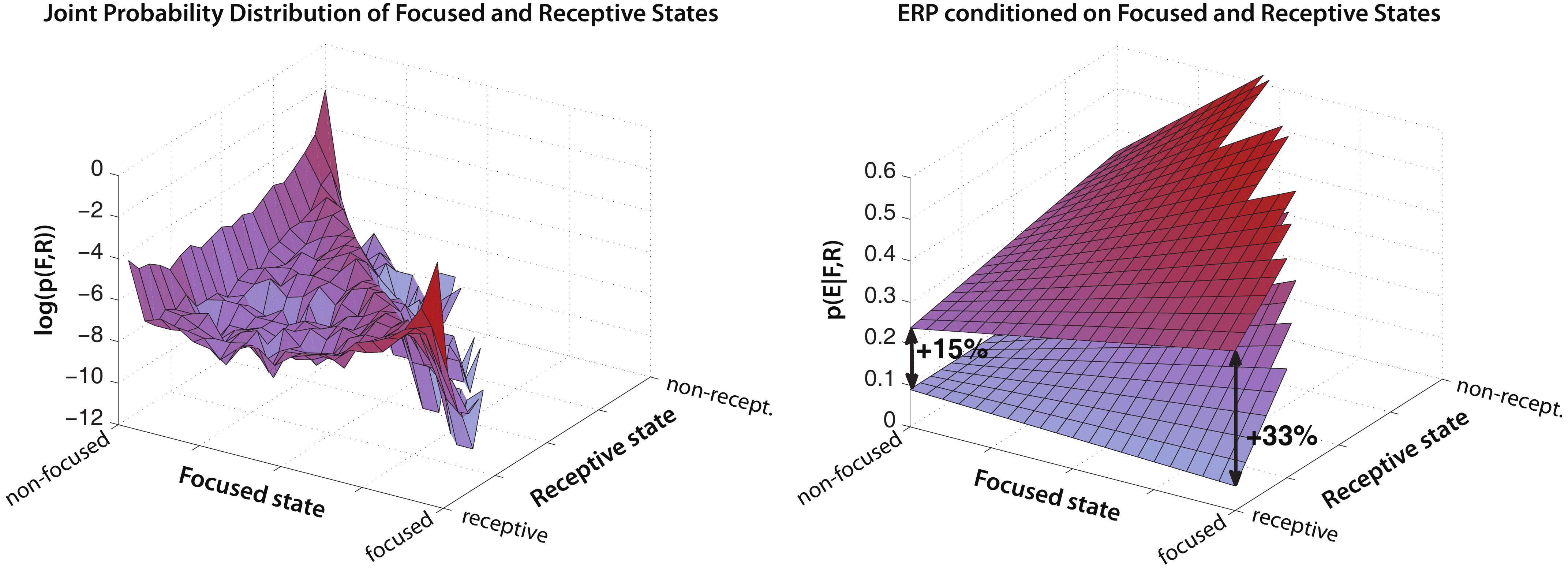}
\caption{
Joint probability distribution of Focused and Receptive states (left) and error repetition probability (ERP) conditioned on engagement states for forgetting (top plane) and no forgetting (bottom plane) (right).
}
\label{fig:engagementstates}
\end{figure}

We particularly investigated the states \emph{Focused} and \emph{Receptive} and analyzed the relation between these states 
by their joint probability distribution revealing that in a fully focused state students are never completely non-receptive.
The plot in Figure~\ref{fig:engagementstates}, left, shows that students can be distracted (non-focused) despite being in a receptive state. 
Figure~\ref{fig:engagementstates}, right, shows the error repetition probability (ERP) conditioned on the two states. It can be seen that in the focused state the offset between the top plane (forgetting) and the bottom plane (no forgetting) is larger than in the non-focused state, i.e., more non-serious errors are committed in the non-focused state.
As expected, the non-receptive state generally causes a higher ERP. 
An interesting observation is the dependency between the age and the engagement state. Our analysis indicates that younger students (with age below the median of $10.34$ years) exhibit a significantly higher probability of being classified as non-receptive ($24.2\%$) and non-focused ($32.5\%$) compared to those above the median ($20.0\%$ and $27.0\%$, respectively). 

The above presented framework has been designed specifically for spelling learning and the training environment \emph{Orthograph}. 
In a subsequent exploratory analysis~\cite{KA13} we investigated to what extent the features and methods are applicable to other learning domains and training systems. 
In particular, we considered the learning of mathematics and the structure of \emph{Calcularis}. While spelling learning is a non-hierarchical process (a word is learned through memorization), learning mathematics is hierarchical (distinct concepts build upon each other). Furthermore, the training environments are significantly different from each other regarding the number of games and their structure (see Section~\ref{sec:domain}).
Therefore, most features that we used in our model for spelling learning cannot be applied to different domains and environments. We analyzed the specific features and suggested a generic feature set for modeling engagement dynamics, considering the categories input behavior, problem statement, problem-solving behavior, performance and environment as listed in~\cite{KA13}.

\subsection{Student Traits}
\label{subsec:traits}
 %
\begin{figure}[t!]
\centering
\includegraphics[trim=0.5cm 11.5cm 0.5cm 11cm, clip=true, width=1.0\columnwidth]{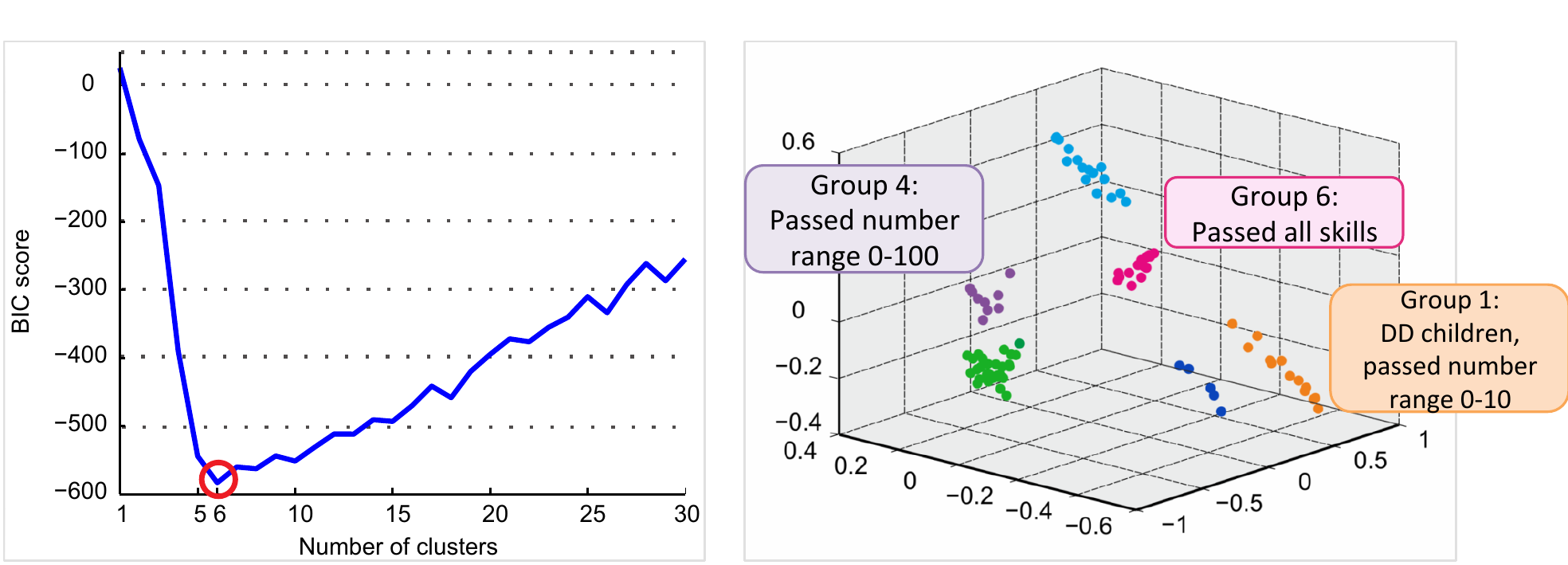}
\caption{
Clustering of student learning patterns, resulting in an optimal BIC for six distinguishable groups (left). The clusters are clearly separable in the three (transformed) dimensions (right).
}
\label{fig:clustering1}
\end{figure}
Besides knowledge and affect, features such as the learner type or learning behavior have a significant impact on the learning outcome. A student trait could for example be a binary classification of whether the student is suffering from a learning disability. In this context, we have developed a screening tool for developmental dyscalculia (DD)~\cite{Kli16b}, which is part of the analytics component and will be discussed in Section~\ref{sec:screener}. Another important trait is learning behavior. We have used clustering approaches to identify multiple subgroups of children with similar learning patterns~\cite{KAE13b}. 
This information can be leveraged for optimizing the student model (training individualization) and to provide visual feedback for domain experts.

Our method consists of two steps: In a first step, we cluster children according to individual learning trajectories. 
The second step consists of a supervised online classification during training, enabling prediction of future performance based on subgroup information only. In particular, we focused on the prediction of 1) long- and short-term training performance (for analysis purposes, to improve adaptation by e.g. minimizing frustration), 2) individual knowledge gaps (to increase the degree of individualization), and 3) external post-test scores (for model validation).
To perform the offline clustering, we first extract a set of features describing cumulative and per skill measures, performance, error behavior and timing (e.g., highest skill reached, number of passed skills, mean answer time per skill) from the recorded log files. The features are then processed and the pairwise dissimilarities between the children are transformed to distances between points in a (higher-dimensional) Euclidean space using kernel transformations. We use the Bayesian Information Criterion (BIC) to determine the optimal number of clusters.
The best BIC was reached with six clusters, visually supported by the clear separability of the transformed data in three dimensions (see Figure~\ref{fig:clustering1}).

The distinct learning patterns of the six clusters were interpreted by experts. For example, students in the best performing cluster (Group 6) passed all skills in the system, while all students in the lowest performing cluster (Group 1) were diagnosed with DD and exhibited difficulties with basic number representation as well as the acquisition of procedural knowledge.

The second step is performed online (i.e. during the training of a student): we classify a student to a particular subgroup in order to be able to predict the future performance based on the subgroup membership. 
 %
The results of the online classification show that after only five training sessions 50\% of the students are correctly classified (chance: 16.6\%). The predictive performance of the model in the areas of interest 1) - 3) mentioned above demonstrates that the accuracy can be significantly increased by taking subgroup information into account. A good prediction accuracy is reached already after a few training sessions and allows us to draw conclusions about short-term performance and knowledge gaps.

\section{Pedagogical Module}
\label{sec:pedagogical}
The pedagogical module is responsible for making optimal teaching decisions based on the current state of a user provided by the student model. The module optimizes the sequence of tasks presented to the student and decides on when to stop teaching a particular skill. 

\begin{figure}[b!]
\centering
\subfigure[Controller in Orthograph]
{\includegraphics[trim=0.6cm 11.5cm 0.5cm 11cm, clip=true, width=0.7\columnwidth]{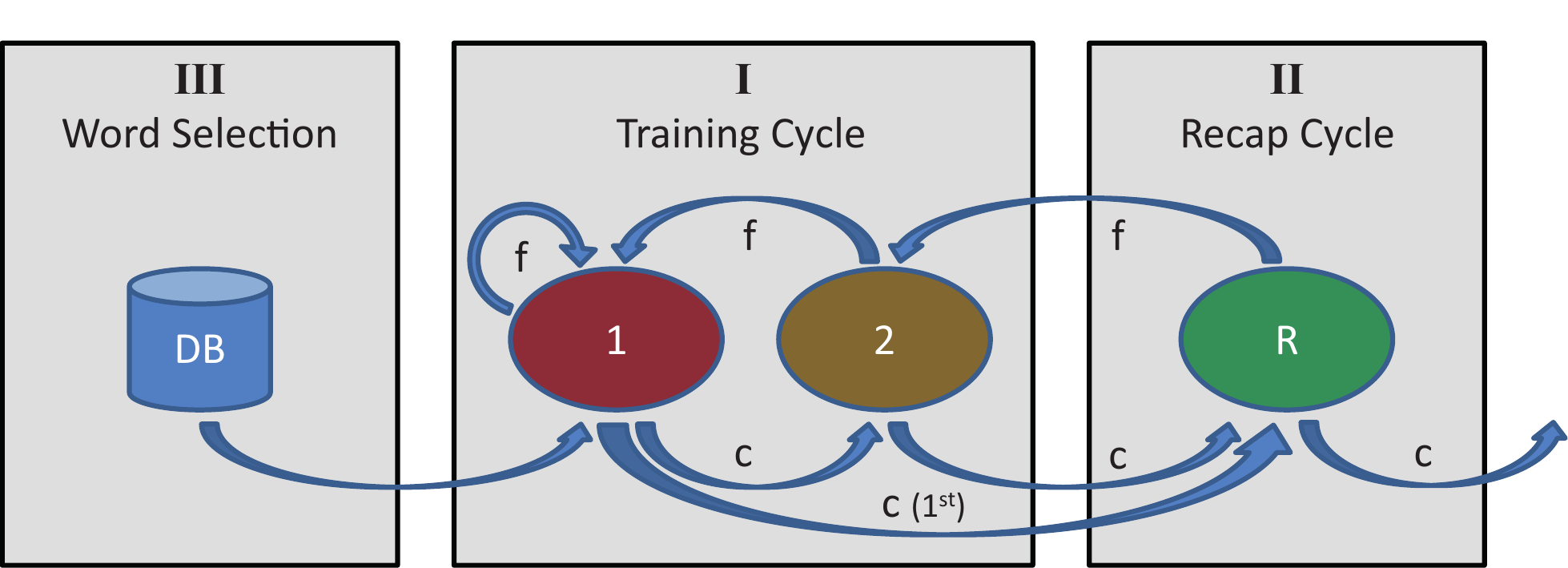}
\label{fig:trainingOrthograph}
}
\subfigure[Controller in Calcularis]
{\includegraphics[trim=0cm 9cm 0cm 9cm, clip=true, width=0.7\columnwidth]{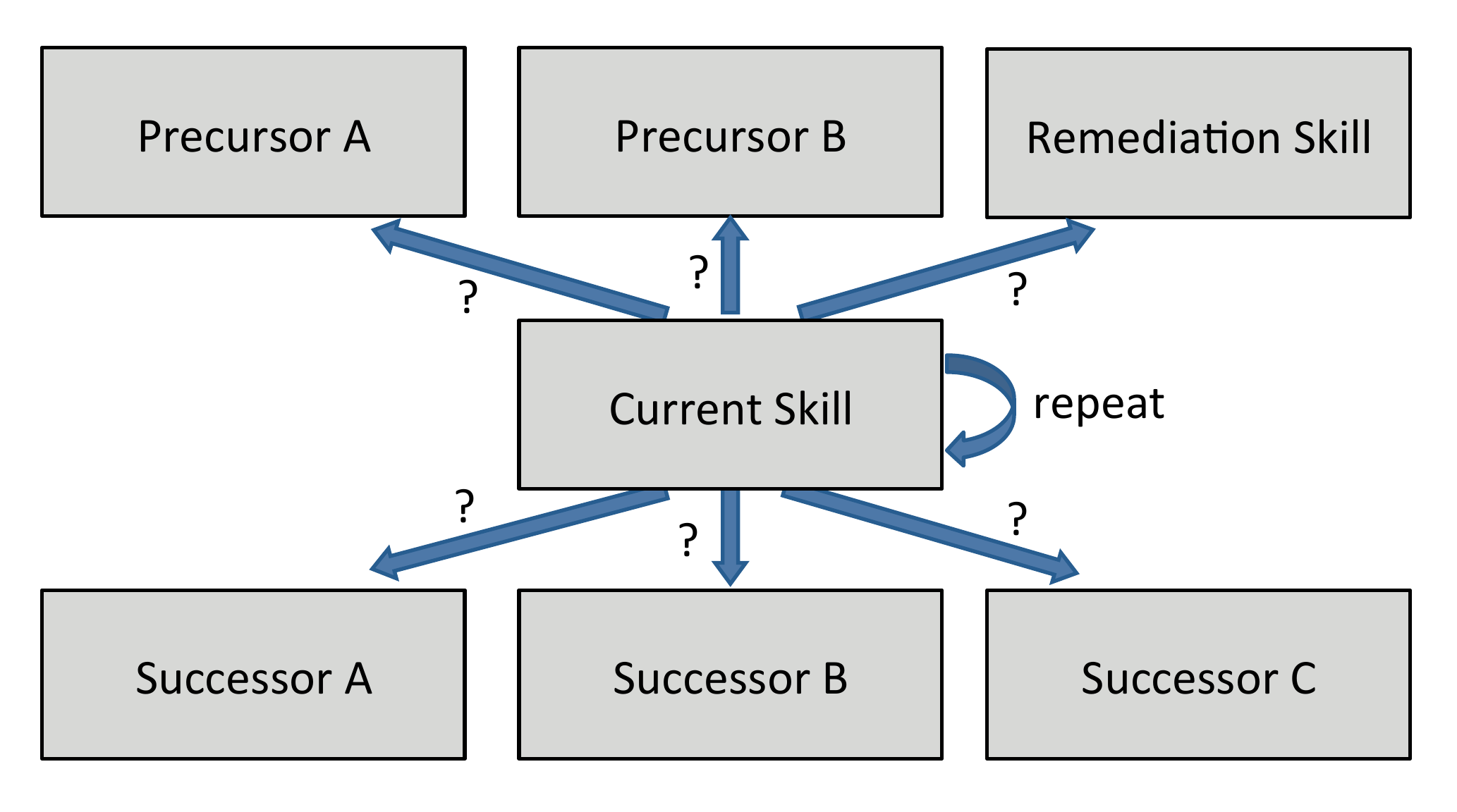}
\label{fig:trainingCalcularis}
}
\caption{
(a) The controller in Orthograph selects the word with the highest error expectation rate. A training and a recap cycle are used until the word has been entered correctly twice. $c$ denotes a correctly spelled word and $f$ a misspelling. $c(1^{st})$ represents a word that was correctly spelled on the first attempt.
(b) The controller in Calcularis navigates a student through the entire skill net. Three actions are possible: Stay, go forward to a successor skill, or go backwards to a precursor or remediation skill. The decision of the next action is based on the posterior probabilities delivered by the student model.
}
\end{figure}

Currently, we only consider the knowledge state of a user in our training environments. 
In Orthograph, our word selection controller selects words with the highest error expectation per letter ratio from the database and thus minimizes the error expectation of the student's input. Generally, words have to be entered correctly twice. If they are entered correctly in the training cycle, the recap cycle is entered. Erroneous inputs lead to a retraining of the word. The training process is illustrated in Figure~\ref{fig:trainingOrthograph}.
\begin{figure*}[ht!]%
\centering 
\includegraphics[trim=0cm 2.75cm 0cm 0cm, clip=true, height=0.55\columnwidth]{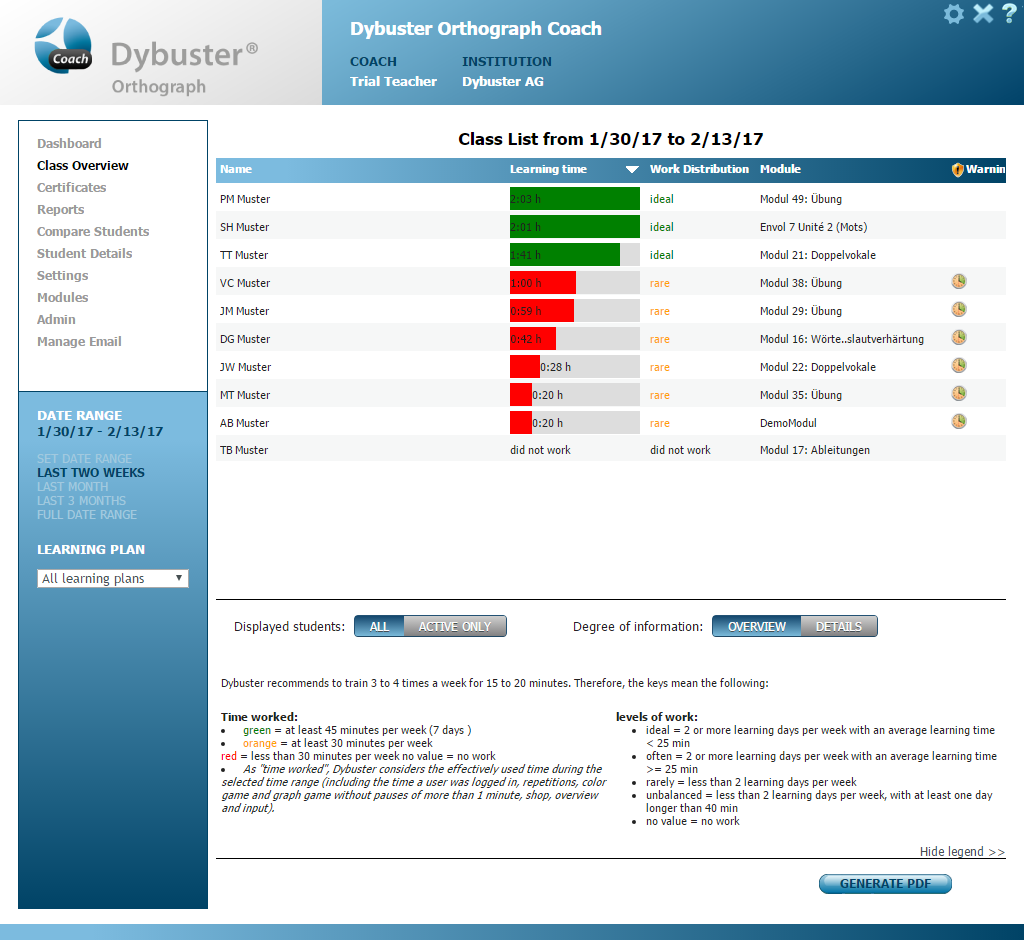}
\hspace{10mm}
\includegraphics[trim=0cm 2.4cm 0cm 0cm, clip=true, height=0.55\columnwidth]{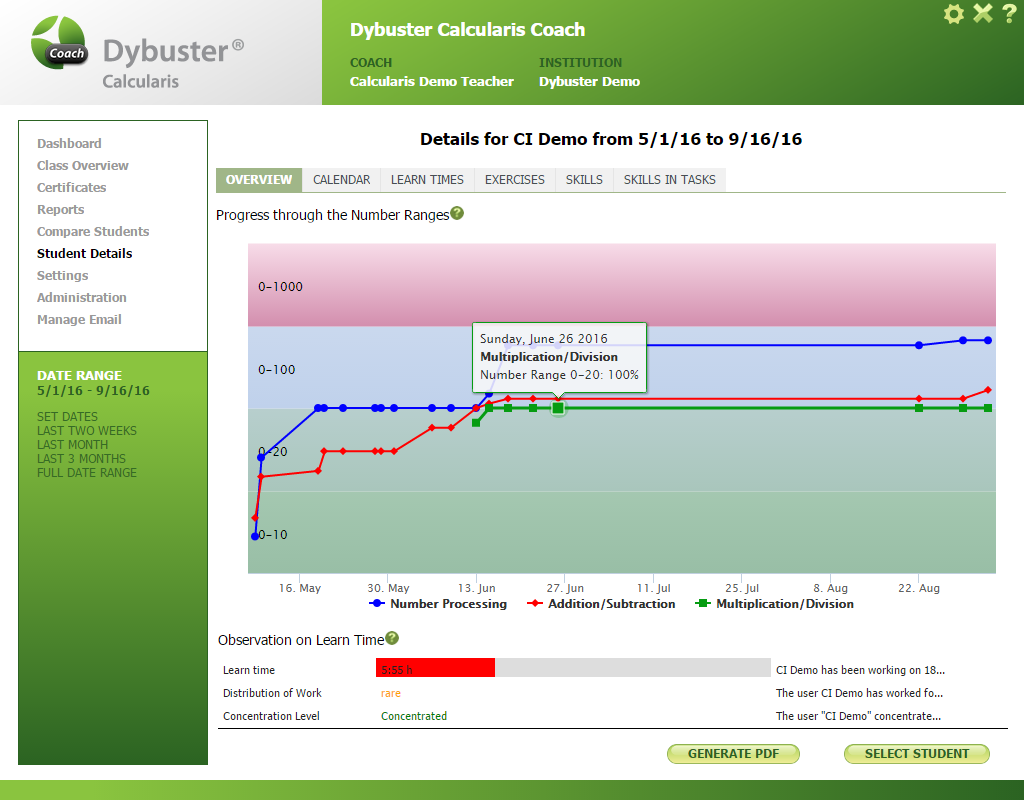}
\\
\vspace{1mm}
\includegraphics[trim=0cm 2.75cm 0cm 0cm, clip=true, height=0.55\columnwidth]{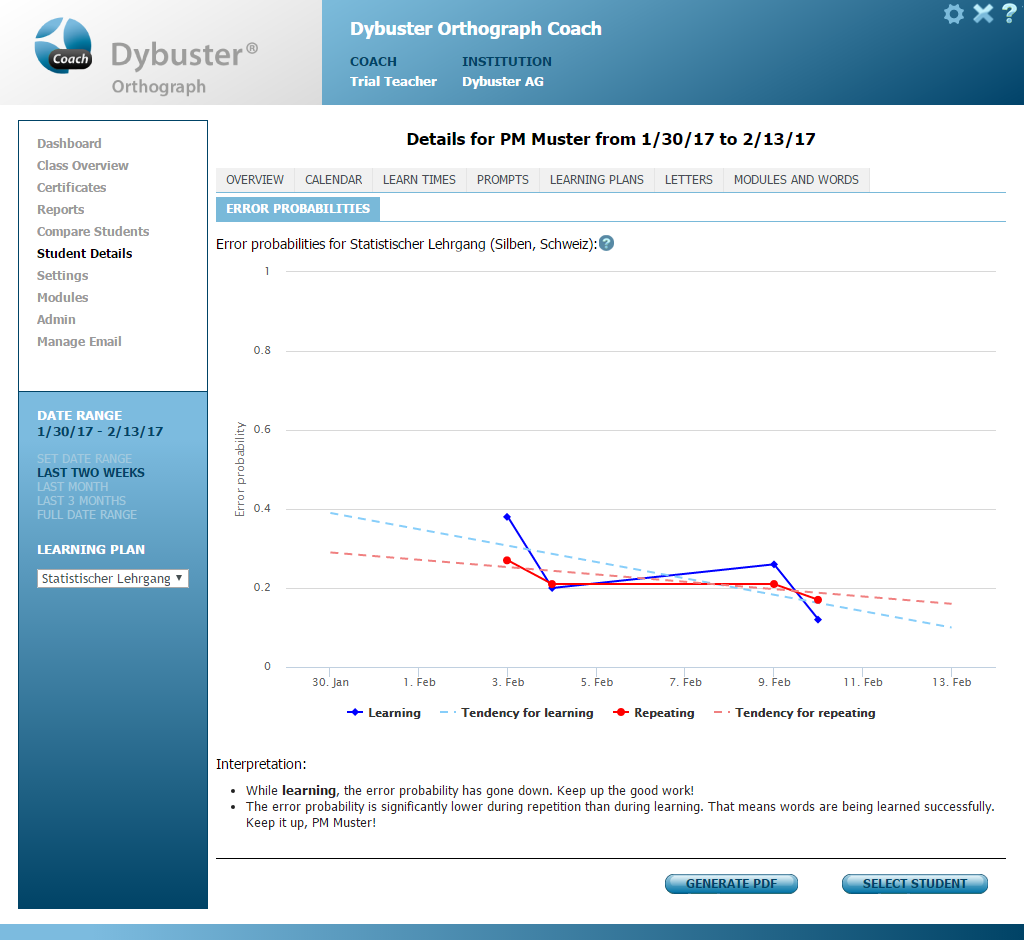}
\hspace{10mm}
\includegraphics[trim=0cm 2.95cm 0cm 0cm, clip=true, height=0.55\columnwidth]{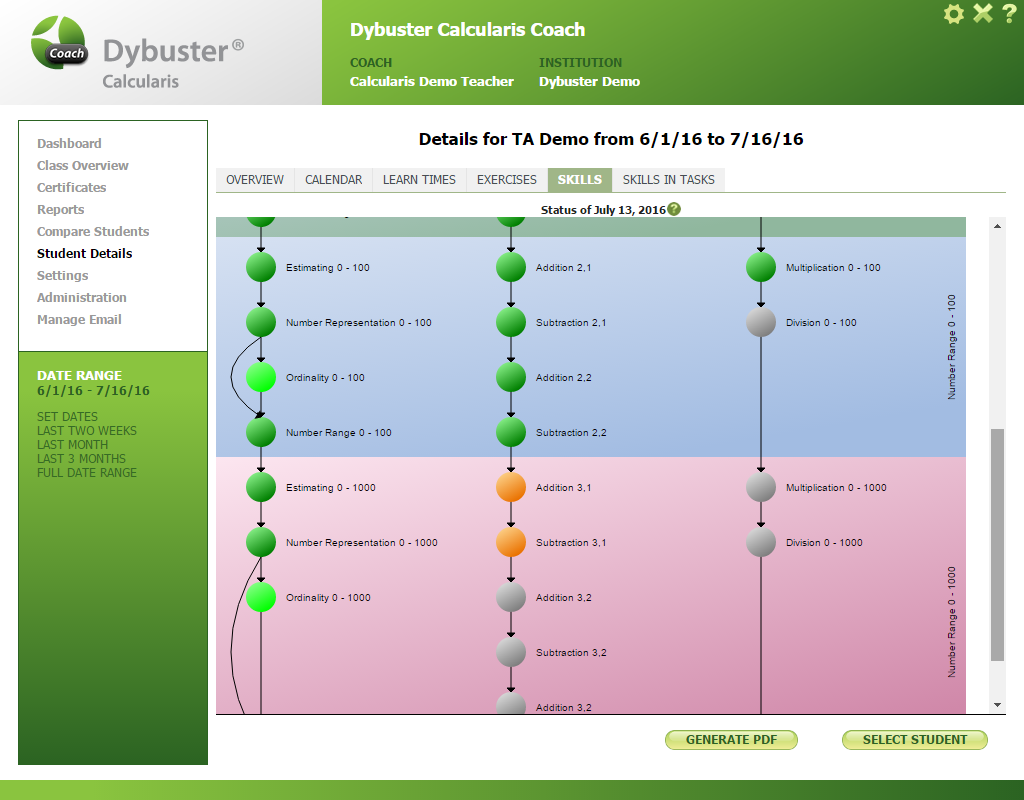}
\caption{
The company Dybuster AG provides detailed analytics and visualization views for teachers in their commercial versions of Orthograph (left) and Calcularis (right). The information supports teachers to inspect learning progress and behavioral patterns of individual children and the class. 
Top left: Overview of all students in a class, showing their learning time and current learning module. 
Bottom left: Plot of the error probability over time for a particular student. The decreasing error probability indicates that words are learnt successfully.
Top right: Overview of the progress through the number ranges. In this example, a progress through the number ranges can be observed over time. 
Bottom right: A graph structure shows the skill status such learned and unlearned for different arithmetic skills.
\textcopyright Image courtesy of Dybuster AG.
}
\label{fig:dybusterCoach}
\end{figure*}

In Calcularis, the controller decides on the next action based on the posterior probabilities delivered by the student model. Three actions are possible as illustrated in Figure~\ref{fig:trainingCalcularis}: stay and continue training of the current skill, go back and train a precursor skill, or go forward and train a successor skill. A rule-based approach is used if there is more than one possibility of precursor and successor skills. When going backwards, it is especially important to consider remediation of typical errors since this allows for better adaptation to the user. Hence, if a typical error is entered the corresponding remediation skill with the lowest probability is displayed. This results in an optimal selection of the next task based on the student's current knowledge state and leads to an individualized training path trough the skill net (Figure~\ref{fig:learningpaths}).

Several factors influence the decision on when to stop teaching a skill (when-to-stop policy). While previous methods typically use policies that are tailored towards the specific task and model, we have proposed a general policy that works for any probabilistic model and can even represent forgetting and identify wheel-spinning students in~\cite{Kae16a}.
When-to-stop policies must take into account several factors to decide to stop teaching a skill. At any point in time a student can be in one of the following situations: 1) A student is mastering a particular skill, but he or she might still make errors when solving tasks associated with that skill. 2) A student is not yet mastering a skill but shows steady progress and is therefore likely to learn the skill given some more repetitions of tasks associated with the skill. 3) A student might be in a situation where his or her current skill set is not sufficient to ever master the new skill required by the task. According to a study by~\cite{Bec13} up to 10\% of students training a particular skill are so called wheel-spinning students. It is vital to detect the different situations accurately in order to avoid over and under practicing of skills as well as frustration of the student. Further it has been shown that over practicing is not necessarily beneficial to the learning outcome. On the contrary, better pedagogical decisions can lead to a decreased effort for the student while retaining the same performance. 

Traditional pedagogical modules employ student model specific when-to-stop policies (e.g. mastery threshold~\cite{Cor94}) that are tailored to the particular task and model. Recently, there has been increasing attention to develop more general policies that only have weak requirements on the student model they work with~\cite{Rol15}. We extended these general policies towards a universal when-to-stop policy that only requires that the student model is able to output the probability of the next task to be correct. The method works for any probabilistic model, and can even represent forgetting and identify wheel-spinning students.
We demonstrated that our new policy is able to handle a large variety of different student models and is robust to noise in the learning process~\cite{Kae16a}: we can identify most wheel-spinning students and stop the training for those particular tasks. We believe that by using generic instructional policies we can strengthen our understanding of how people are learning, as it disentangles effects from the accuracy of the student models and effects from the pedagogical module.

\section{Analytics and Visualization Engine}


Our gamified learning platforms have logged every key stroke, task prompt, and time information from student trainings over multiple weeks, months, and years. This vast amount of data renders manual inspection by human experts impractical. 
For teachers, statistics and information to inspect the learning progress and behavioral patterns of individual children are provided by the commercial versions of Orthograph and Calcularis. Screenshots of the softwares are shown in Figure~\ref{fig:dybusterCoach}, illustrating for example the time spent learning, learning progress over time and current skill level.
For domain experts and researchers, our developed methods aim at providing support to successfully differentiate between important patterns in the data and to reason about the semantic meanings of such knowledge structures in the context of computer-aided learning. 
Research questions may include the identification of individual learner types and patterns or understanding the interaction with the learning environment. 
%
In the following we will discuss two of our applications, the temporal clustering and the screening tool for developmental dyscalculia.
\subsection{Temporally Coherent Clustering}
\label{sec:evolutionary}

\begin{figure}[t!]
\centering
\includegraphics[trim=5.125cm 4.25cm 5.2cm 7.2cm, clip=true, width=1.0\columnwidth]{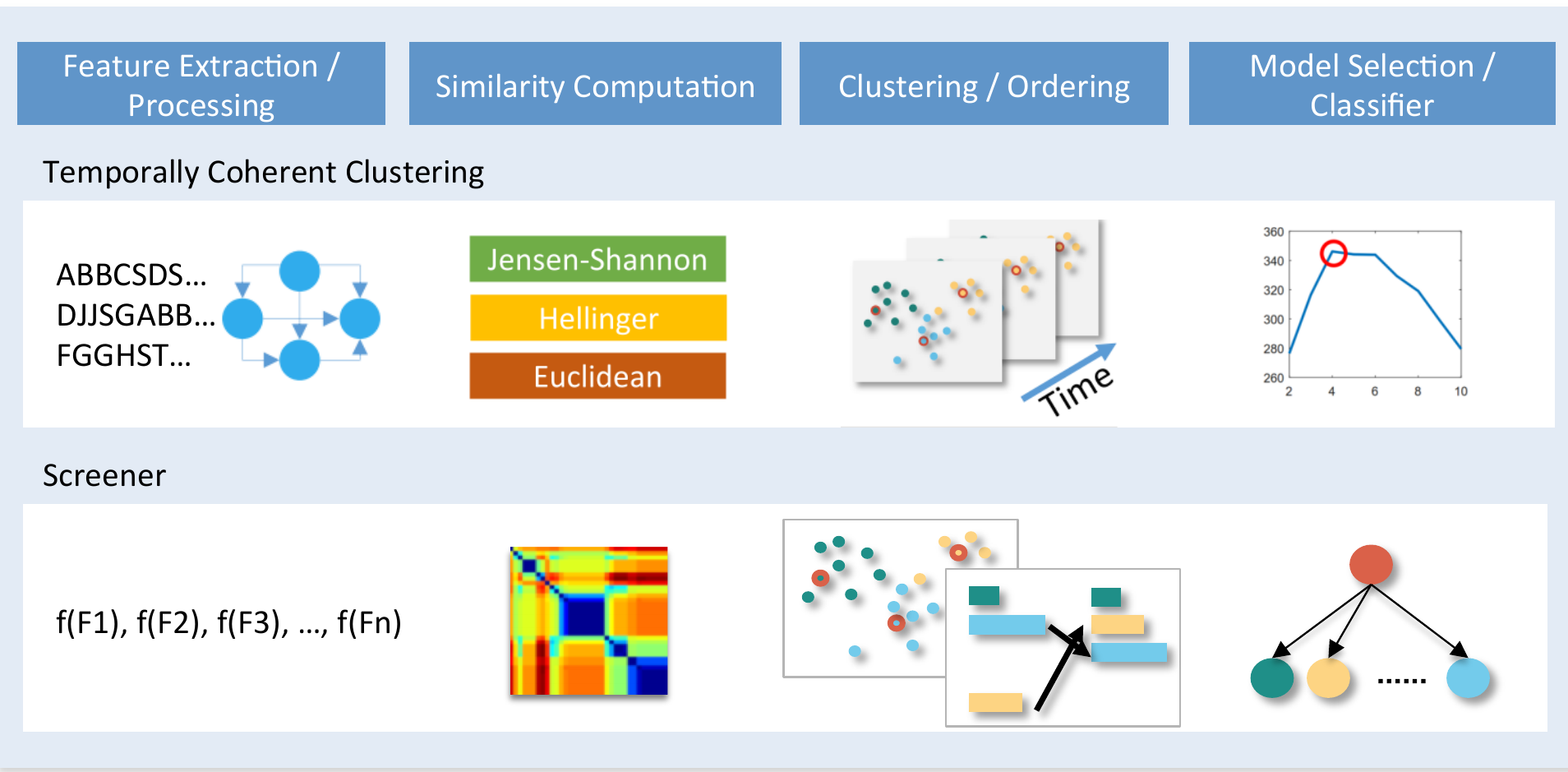}
\caption{Our pipeline showing the processing steps for the temporally coherent clustering and the screening tool for developmental dyscalculia.}
\label{fig:pipeline}
\end{figure}

The clustering of learner types discussed in Section~\ref{subsec:traits} is not only 
used by the knowledge engine but serves as a reasoning tool for experts as well. The clusters reveal similar students and allow for short- and long-term prediction of the student's knowledge and skill level. 
In contrast to this analysis, a temporal analysis of student clusters would allow experts to identify how interaction patterns change over time and how similar group of students evolve. We have developed a pipeline (Figure~\ref{fig:pipeline}) to capture such relevant cluster evolution effects that can also be used as a black box for data coming from any intelligent tutoring system~\cite{Kli16}. 

We use an evolutionary clustering approach that is less sensitive to noise than static approaches. The temporal smoothing improves cluster stability significantly and allows for a better analysis of the clusters. 
In a first step, we extract specific action sequences from the ITS that we want to study.
We transform the sequences into an aggregated representation using Markov Chain models to remove noise. 

\begin{figure}[t!]
\centering
\includegraphics[trim=0cm 7cm 0cm 7cm, clip=true, height=0.28\columnwidth]{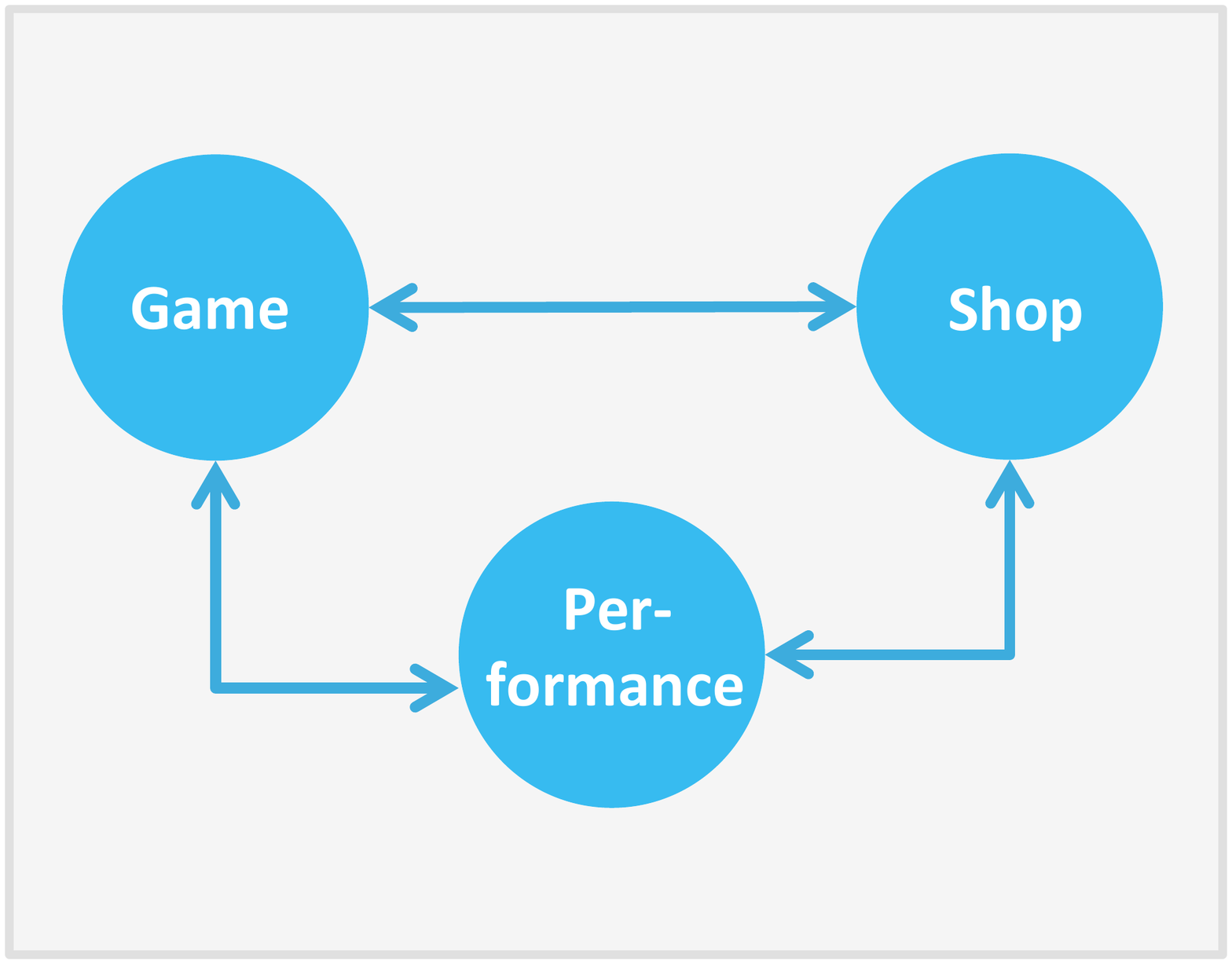}
\includegraphics[trim=0cm 1.75cm 12cm 0cm, clip=true, height=0.28\columnwidth]{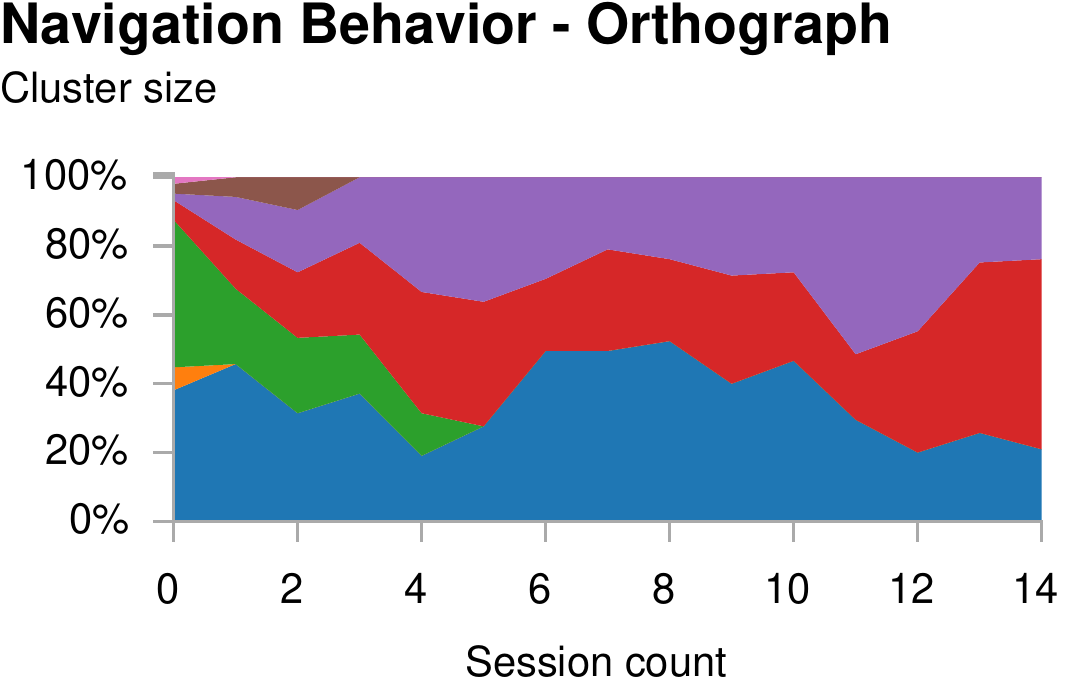}
\includegraphics[trim=0cm 7cm 0cm 7cm, clip=true, height=0.28\columnwidth]{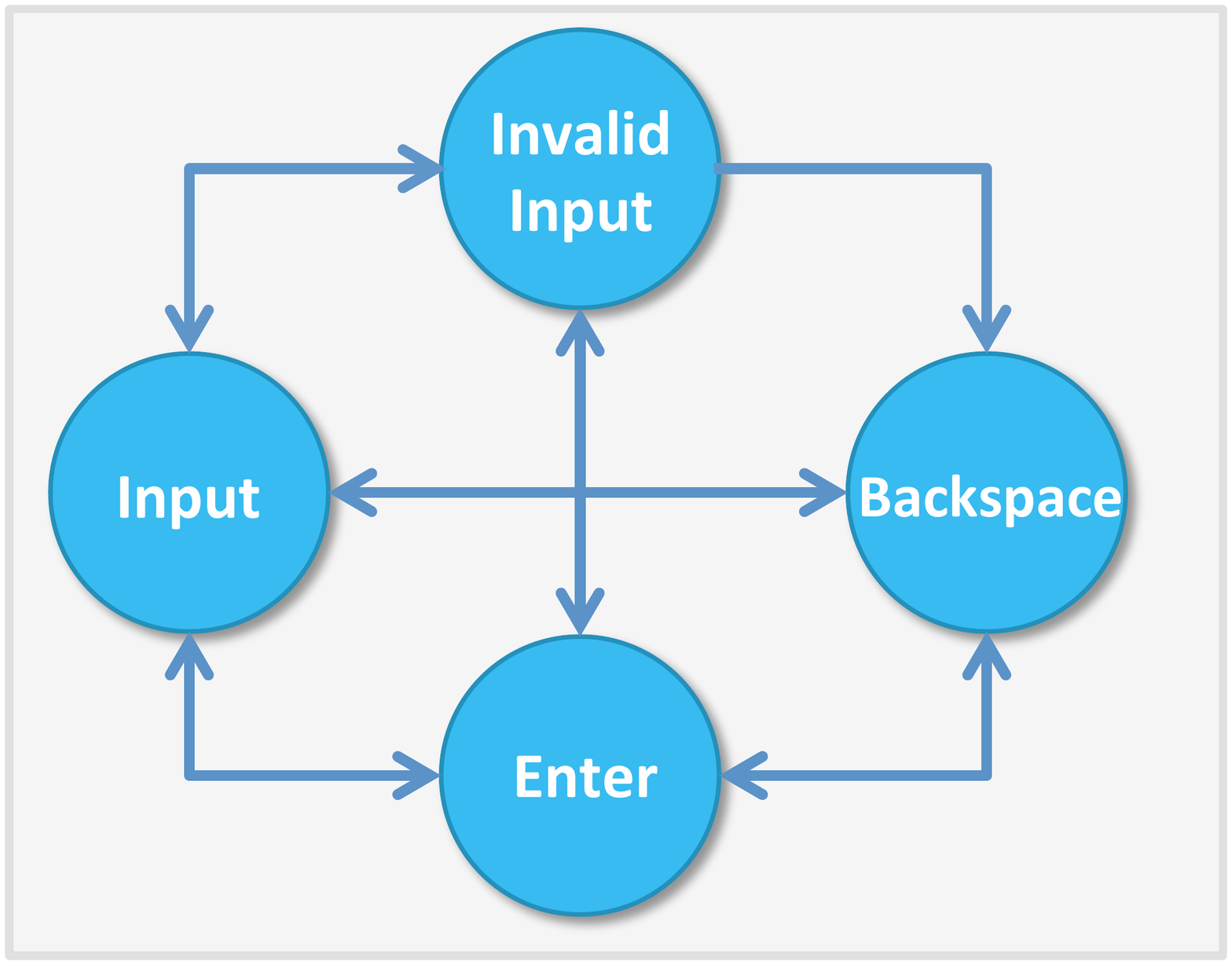}
\includegraphics[trim=0cm 1.75cm 12cm 0cm, clip=true, height=0.28\columnwidth]{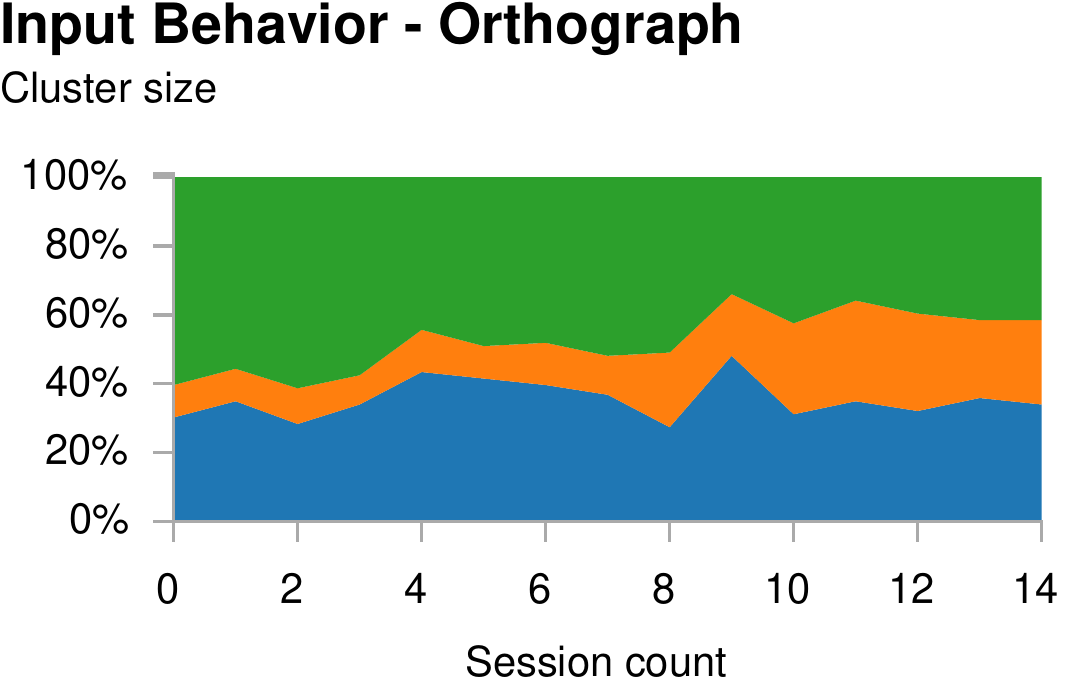}
\caption{Left: Markov Chains capturing the navigation (top) and input (bottom) behavior in our spelling environment \emph{Orthograph}. Right: The corresponding color-coded student clusters over time.}
\label{fig:clustering2results}
\end{figure}

Figure~\ref{fig:clustering2results} shows these models for the navigation and input behavior in Orthograph.
The three navigation states - \emph{Game}, \emph{Shop} and \emph{Performance} - represent game play, spending the collected points during game play in the shop, and analyzing the progress in a current word module. The input behavior in \emph{Orthograph} is represented by four states, where \emph{Input} / \emph{Invalid Input} refer to entering a correct / incorrect letter, \emph{Backspace} indicates the correction of a single letter, and \emph{Enter} corresponds to entering the complete word.

The edges between two states define the pairwise transition probabilities of actions. 
We compute pairwise similarities between two Markov Chains, smooth the similarity matrices temporally, and then perform clustering.
The amount of smoothing is chosen adaptively such that the system relies more heavily on previous observations if the estimated noise is high, and more on current observations if there is a large amount of new information (such as students showing a novel behavior). Finally, the smoothed similarity matrices are clustered with K-Means.
%
The visualization of the clusters has proven to be a powerful tool for domain experts and researchers to reveal behavioral patterns of students but also to detect potential deficiencies of the learning environment. To exemplify this, we show the clearly distinguishable student clusters (color-coded) over time for the input and navigation behavior (Figure~\ref{fig:clustering2results}, right).
For the navigation behavior, we observe three stable clusters (purple, red and blue) after a few initial sessions. Children in the blue cluster are very focused on the training as they spend 82\% of their time in the game, while children in the red cluster spend a significant time of their time (34\%) in the shop. The students in the purple cluster tend to navigate frequently between the states.
For the input behavior, we observe a similar pattern: we again obtain three stable clusters over time. The green cluster consists of focused students who consistently produce a low percentage of invalid inputs, while the error rate in the orange cluster is high, and the performance of students in the blue cluster lies in-between.

\subsection{Non-intrusive Screening for DD}
\label{sec:screener}

As a stand-alone test we have developed a screening tool that non-intrusively screens a student for developmental dyscalculia (DD) using the games of \emph{Calcularis}. The current prototype currently serves as an analytics tool for teachers and experts. We envision to integrate the screener seamlessly into our architecture, i.e., as a module of the student trait detector, and hence to run it directly in the background of the learning system. Due to these beneficial properties, such a model may represent the first step towards the highly desired universal and inexpensive screening of children at early years in school. 

Our pipeline (Figure~\ref{fig:pipeline})~\cite{Kli16b} can be used as a black box for various game and learning environments. We validated our model for the case of detecting children that suffer from DD while they are playing with the games in \emph{Calcularis}.
We extracted a few hundred features from the log file data of the trainings and clustered similar features into groups. We then selected the most representative feature per group for the classification.
%
 %
%
Based on the selected features, we use a probabilistic model that adapts the test duration to the individual student. The classification task is solved using an adapted Naive Bayes model, which assumes conditional independence of all the features. To determine the optimal ordering of the tasks in the test, we compute the amount of information in each feature using an unpaired t-test and ordering them according to their p-value, starting with the feature that has the smallest one.

\begin{figure}[t]
\centering
\subfigure[Selected features and their ordering.]
{
\includegraphics[trim=0cm 7cm 0cm 7cm, clip=true, width=0.8\columnwidth]{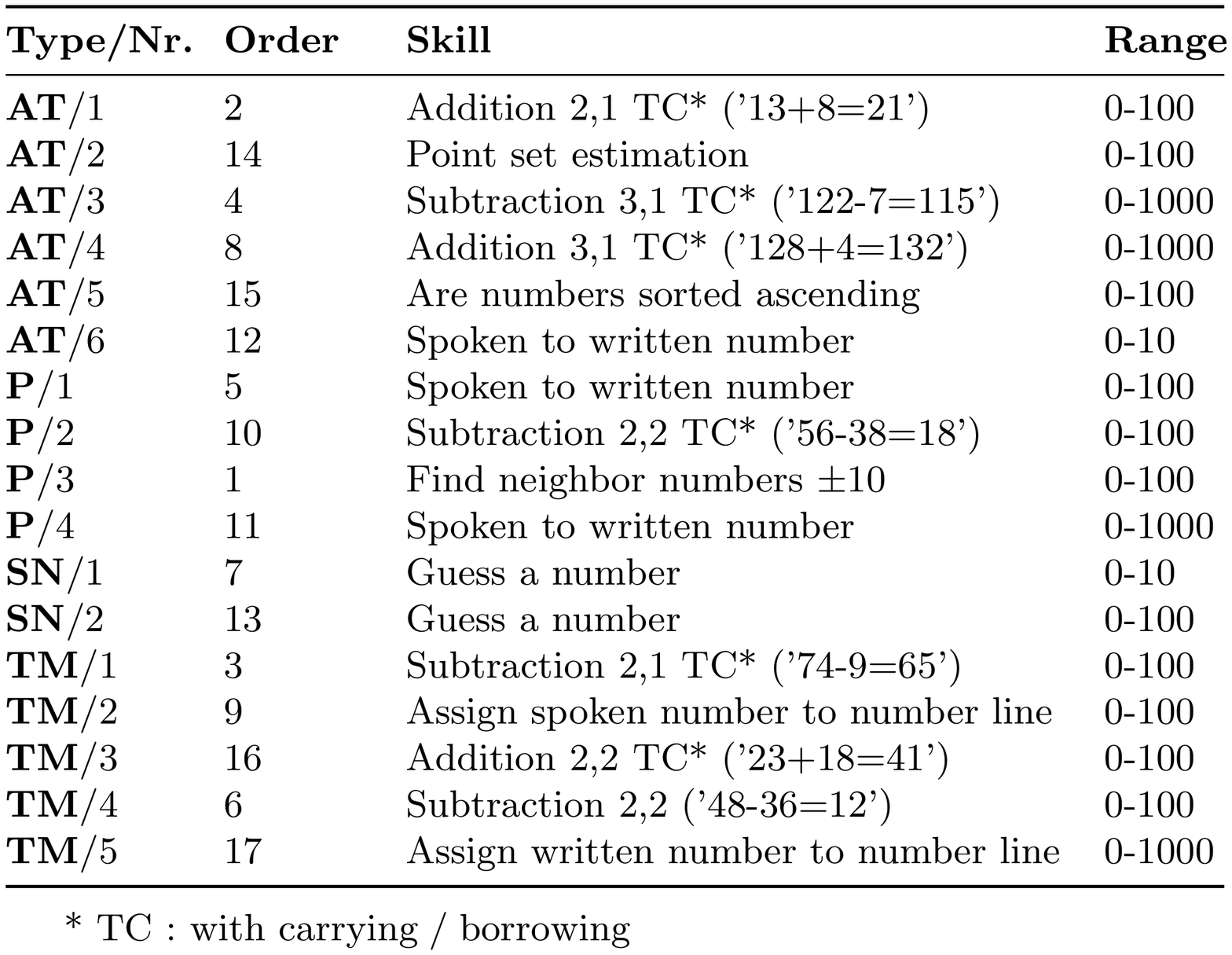}
\label{fig:featuretable}
}
\subfigure[Criterion-related validity.]
{
\includegraphics[trim=0cm 0cm 0cm 0cm, clip=true, width=0.8\columnwidth]{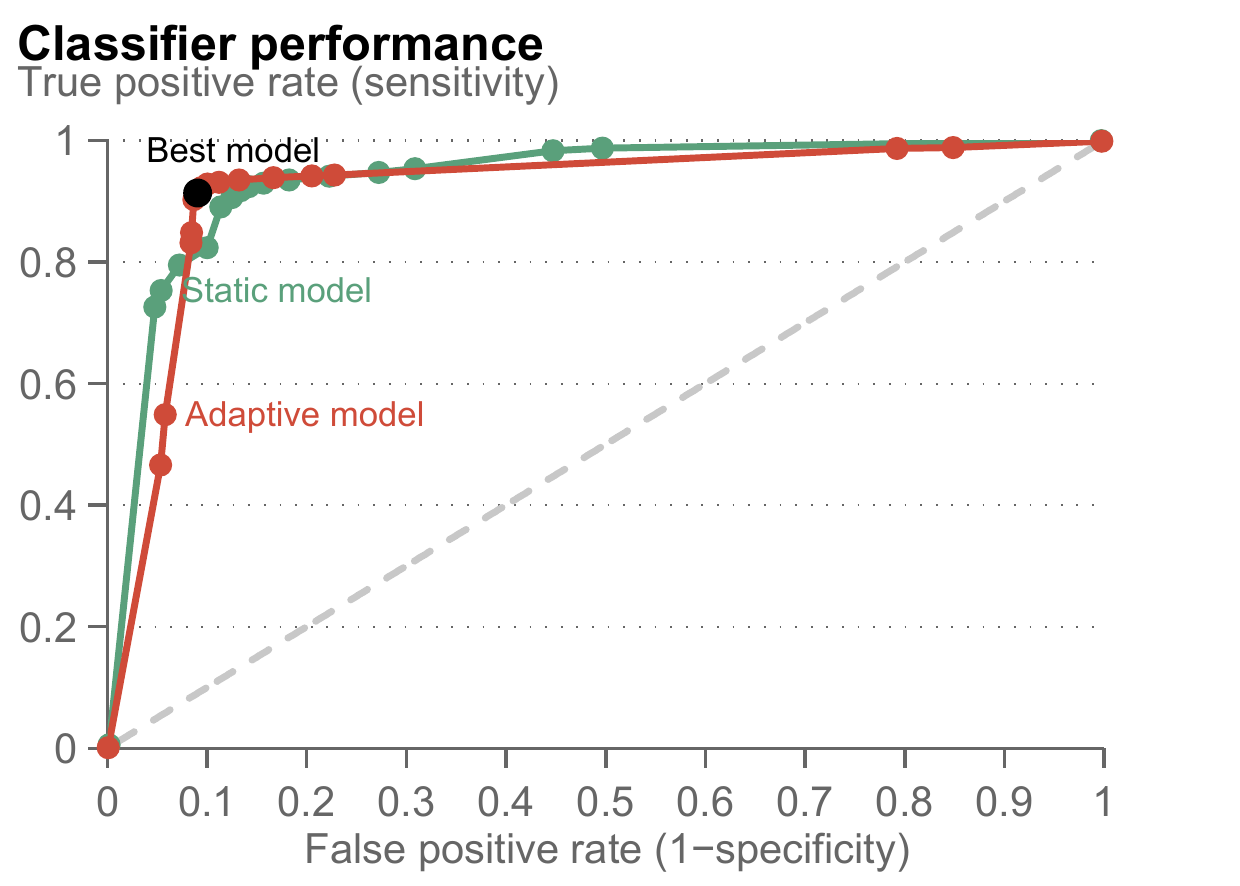}
\label{fig:reliability}
}
\caption{We identified a set of features (a) for screening children for DD and reached a specificity and sensitivity of 0.91 with our best model (b).}
\end{figure}

$17$ features were automatically selected based on the recorded data alone (Figure~\ref{fig:featuretable}). 
The features agree well with findings in previous work on DD. For example, the performance value P/3 (ratio of the correctly solved tasks) and answer time AT/5 capture the deficits in number comparison that are shown by children with DD, whereas number processing skills are captured by AT/2, AT/6, P/1, and P/4. Typical mistakes (TM) and difficulties in problem solving strategies (SN) can be mapped onto findings in DD as well.
Moreover, the selected features correspond well with the type of tasks used in standardized tests for DD such as counting, number comparison, number representation and simple arithmetical tasks. 
Interestingly, the screener includes some features such as typical mistakes and problem solving strategies that are not captured by traditional paper tests for identifying DD.

The test duration is adapted to the individual child as we stop the test once we observe that a new feature does not change the classifier's current believe of the group label. 
On average, our adaptive screener classifies a child as being at risk for DD or not DD after only $11$ minutes - this is notably shorter than screener durations reported for digital and non-digital screeners in previous work. Already after five test minutes, $40\%$ of the children are classified. 
Our best probabilistic classifier exhibits a high sensitivity and specificity of 0.91 (Figure~\ref{fig:reliability}).

\section{Lessons Leaned and Next Generation Educational Games}



Our design decision to store each key stroke of a student has proven to be beneficial. We have a large and powerful data set at hand that allows to analyze and reveal complex behavioral patterns. Moreover, all our methods are based on the same set of data, which means that our processing pipeline can be used as a black box in many of the modules where feature extraction and processing is relevant. 

While our evaluation of each module has shown very good results, one important future work is to understand the interplay of the modules and to evaluate the whole architecture. 
In general, proper scientific evaluations of such complex systems are largely missing in ITS. There is a tremendous need for fast, semi-automatic randomized trials in order to have fast turn-around time when testing different variants of an algorithm (e.g. different controllers). Today a lot of the research is still based on relatively small sample sizes in controlled manually administered studies.


Due to the recent trend of extending traditional one-to-one learning therapy in schools and the transformation of paper-and-pen exercises to digital learning environments (such as intelligent tutoring systems, educational games, learning simulations, and massive open online courses (MOOCs) that produce high volumes of data, researchers will have access to large training data sets of students. Such data is extremely powerful as it enables statistical interpretations that can be used to develop novel data-driven models for educational games. The non-uniformity and inhomogeneity of the students and data play an important role as the data covers a wide range of student characteristics and learning patterns. Leveraging these properties will allow us to develop student models and expert analytics tools with higher accuracy and higher predictive power. 
The presented screener of children for the learning disability dyscaculia is one example that shows the potential of identifying such properties non-intrusively while a student is interacting with the system. 

Another factor that, in our opinion, requires more attention is to reliably capture the student's affect. Previous work explored the use of bio-sensors and video data to estimate if a student is engaged or bored. 
We envision, however, an approach that works with log file data only and thus renders intrusive and distracting hardware setups obsolete. The results of our preliminary study on engagement modeling shows great potential for such techniques.
In addition, models have to be developed and evaluated that offer an optimal intervention based on the identified affect. 
Current approaches typically work with integrated animations, digital assistants and game play elements such as a leader board or point-and-shop system.
We believe that further motivational elements need to be explored, such as collaborative learning, more complex game-play and storytelling, or improved graphics and AR.


%
%
We believe that storytelling elements should be explored. In our opinion, wrapping a learning system into a more interesting environment and story has great potential for increasing the gameplay experience and fostering student motivation. Such a story wrapper can develop its full potential if natural language processing is included as this enables a new level of user interaction with the game or system.
Moreover, for our future learning environment we envision the use of Augmented Reality. AR may offer versatile advantages. The combination of visual, auditory and haptic cues will create active learning experiences and engage students to explore and interact. The additional sensory cue will enrich the data set at hand and improve the computational models eventually.
Our presented architecture can be easily extended with such concepts by complementing the current student model. 

Being able to create a virtual representation of a user, i.e., understanding the user and her complex behavior and to react accordingly, has undoubtedly a great potential in other applications than ITS as well. Especially the AI components in video games, recommender systems in media applications, and interactive virtual reality environments could benefit from personalized and adaptive content. For example, today's video games often include only rather simple heuristics and game adaptivity based on the observed user actions (such as dynamic difficulty adjustments~\cite{Bir05}). Ongoing research in this field aims at leveraging machine learning to improve the game play. For example, player types were automatically identified in Minecraft based on log data of the user's interaction with the game \cite{2015-fdg-player-behavior}, or clustering has been used to analyze user behavior in click-stream data~\cite{Wan16}.
We argue that sophisticated user models that consider the knowledge, traits and affect have the potential to transform and improve existing applications even beyond educational games, and we hope that this article displayed the potential of such an interplay.

\section*{Acknowledgment}
This work was supported by ETH Research Grant ETH-2313-2. 
We wish to thank our past co-authors who contributed to the presented models: GM. Baschera, AG. Busetto, M. Kast, J. Kohn, K. Kucian, M. Meyer, L. Jaencke, AG. Schwing, C. V\"ogeli, M. von Aster.
Thanks to Dybuster AG for their industry perspective.
\vspace{2cm}
\ifCLASSOPTIONcaptionsoff
  \newpage
\fi



\bibliographystyle{IEEEtran}
\bibliography{related}

\begin{thebibliography}{10}
\providecommand{\url}[1]{#1}
\csname url@samestyle\endcsname
\providecommand{\newblock}{\relax}
\providecommand{\bibinfo}[2]{#2}
\providecommand{\BIBentrySTDinterwordspacing}{\spaceskip=0pt\relax}
\providecommand{\BIBentryALTinterwordstretchfactor}{4}
\providecommand{\BIBentryALTinterwordspacing}{\spaceskip=\fontdimen2\font plus
\BIBentryALTinterwordstretchfactor\fontdimen3\font minus
  \fontdimen4\font\relax}
\providecommand{\BIBforeignlanguage}[2]{{%
\expandafter\ifx\csname l@#1\endcsname\relax
\typeout{** WARNING: IEEEtran.bst: No hyphenation pattern has been}%
\typeout{** loaded for the language `#1'. Using the pattern for}%
\typeout{** the default language instead.}%
\else
\language=\csname l@#1\endcsname
\fi
#2}}
\providecommand{\BIBdecl}{\relax}
\BIBdecl

\bibitem{Csi90}
M.~Csikszentmihalyi, \emph{Flow: The Psychology of Optimal Experience}.\hskip
  1em plus 0.5em minus 0.4em\relax Harper-Row, 1990.

\bibitem{KM07}
M.~Kast, M.~Meyer, C.~V\"{o}geli, M.~Gross, and L.~J\"{a}ncke, ``Computer-based
  {M}ultisensory {L}earning in {C}hildren with {D}evelopmental {D}yslexia.''
  \emph{Restorative Neurology and Neuroscience}, vol.~25, no. 3-4, pp.
  355--369, 2007.

\bibitem{Bas10b}
M.~Kast, G.-M. Baschera, M.~Gross, L.~J\"{a}ncke, and M.~Meyer,
  ``Computer-based learning of spelling skills in children with and without
  dyslexia,'' \emph{Annals of Dyslexia}, 2011.

\bibitem{baschera2010poisson}
G.-M. Baschera and M.~Gross, ``Poisson-based inference for perturbation models
  in adaptive spelling training,'' \emph{Journal of Artificial Intelligence in
  Education}, vol.~20, no.~4, pp. 333--360, 2010.

\bibitem{KA12}
T.~K{\"a}ser, A.~G. Busetto, G.-M. Baschera, J.~Kohn, K.~Kucian, M.~von Aster,
  and M.~Gross, ``Modelling and optimizing the process of learning
  mathematics,'' in \emph{Proc. Intelligent Tutoring Systems}, 2012, pp.
  389--398.

\bibitem{KA13b}
T.~K\"{a}ser, G.-M. Baschera, J.~Kohn, K.~Kucian, V.~Richtmann, U.~Grond,
  M.~Gross, and M.~von Aster, ``{Design and evaluation of the computer-based
  training program Calcularis for enhancing numerical cognition},''
  \emph{Front. Psychol.}, 2013.

\bibitem{KA13d}
T.~K{\"a}ser, A.~G. Busetto, B.~Solenthaler, G.-M. Baschera, J.~Kohn,
  K.~Kucian, M.~von Aster, and M.~Gross, ``{Modelling and Optimizing
  Mathematics Learning in Children},'' \emph{IJAIED}, 2013.

\bibitem{Sha08}
L.~Shams and A.~R. Seitz, ``Benefits of multisensory learning,'' \emph{Trends
  in cognitive sciences}, vol.~12, no.~11, pp. 411--417, 2008.

\bibitem{von2007number}
M.~G. Von~Aster and R.~S. Shalev, ``Number development and developmental
  dyscalculia,'' \emph{Developmental Medicine \& Child Neurology}, vol.~49,
  no.~11, pp. 868--873, 2007.

\bibitem{Gro07}
M.~Gross and C.~Voegeli, ``A multimedia framework for effective language
  training,'' \emph{Computers \& Graphics}, vol.~31, no.~5, pp. 761--777, 2007.

\bibitem{Sch12}
A.~G. Schwing, T.~Hazan, M.~Pollefeys, and R.~Urtasun, ``{Efficient Structured
  Prediction with Latent Variables for General Graphical Models},'' in
  \emph{Proc. ICML}, 2012.

\bibitem{kae14a}
T.~K{\"a}ser, A.~G. Schwing, T.~Hazan, and M.~Gross, ``{Computational Education
  using Latent Structured Prediction},'' in \emph{Proceedings of Artificial
  Intelligence and Statistics (AISTATS)}, 2014, pp. 540--548.

\bibitem{kae14b}
T.~K{\"a}ser, S.~Klingler, A.~G. Schwing, and M.~Gross, ``{Beyond Knowledge
  Tracing: Modeling Skill Topologies with Bayesian Networks},'' in \emph{Proc.
  ITS}, 2014, pp. 188--198.

\bibitem{BA11}
G.~M. Baschera, A.~G. Busetto, S.~Klingler, J.~Buhmann, and M.~Gross,
  ``Modeling {E}ngagement {D}ynamics in {S}pelling {L}earning,'' in \emph{Proc.
  AIED}, 2011, pp. 31--38.

\bibitem{KA13}
T.~K{\"a}ser, G.-M. Baschera, A.~G. Busetto, S.~Klingler, B.~Solenthaler, J.~M.
  Buhmann, and M.~Gross, ``{Towards a Framework for Modelling Engagement
  Dynamics in Multiple Learning Domains},'' \emph{IJAIED: Best of AIED 2011 -
  Part 2}, 2012.

\bibitem{Kli16b}
S.~Klingler, T.~K{\"a}ser, A.~Busetto, B.~Solenthaler, J.~Kohn, M.~von Aster,
  and M.~Gross, ``{Stealth Assessment in ITS - A Study for Developmental
  Dyscalculia},'' in \emph{Proceedings of Intelligent Tutoring Systems (ITS)},
  2016.

\bibitem{KAE13b}
T.~K{\"a}ser, A.~G. Busetto, B.~Solenthaler, J.~Kohn, M.~von Aster, and
  M.~Gross, ``Cluster-based prediction of mathematical learning patterns,'' in
  \emph{Proc. AIED}, 2013.

\bibitem{Kae16a}
T.~K\"{a}ser, S.~Klingler, and M.~Gross, ``When to stop?: Towards universal
  instructional policies,'' in \emph{Proceedings of the Sixth International
  Conference on Learning Analytics \& Knowledge}, 2016, pp. 289--298.

\bibitem{Bec13}
J.~E. Beck and Y.~Gong, ``Wheel-spinning: Students who fail to master a
  skill,'' in \emph{International Conference on Artificial Intelligence in
  Education}.\hskip 1em plus 0.5em minus 0.4em\relax Springer, 2013, pp.
  431--440.

\bibitem{Cor94}
A.~T. Corbett and J.~R. Anderson, ``Knowledge tracing: Modeling the acquisition
  of procedural knowledge,'' \emph{User modeling and user-adapted interaction},
  vol.~4, no.~4, pp. 253--278, 1994.

\bibitem{Rol15}
J.~Rollinson and E.~Brunskill, ``From predictive models to instructional
  policies.'' \emph{International Conference on Educational Data Mining}, 2015.

\bibitem{Kli16}
S.~Klingler, T.~K{\"a}ser, B.~Solenthaler, and M.~Gross, ``{Temporally Coherent
  Clustering of Student Data},'' in \emph{Proc. Educational Data Mining}, 2016.

\bibitem{Bir05}
D.~Birlew, \emph{ResidentEvil4: Official Strategy Guide}.\hskip 1em plus 0.5em
  minus 0.4em\relax BradyGames, 2005.

\bibitem{2015-fdg-player-behavior}
S.~Muller, M.~Kapadia, S.~Frey, S.~Klingler, R.~Mann, B.~Solenthaler,
  R.~Sumner, and M.~Gross, ``Statistical analysis of player behavior in
  minecraft,'' in \emph{Foundations of Digital Games}, 2015.

\bibitem{Wan16}
G.~Wang, X.~Zhang, S.~Tang, H.~Zheng, and B.~Y. Zhao, ``Unsupervised
  clickstream clustering for user behavior analysis,'' in \emph{Human Factors
  in Computing Systems}, 2016, pp. 225--236.

\end{thebibliography}

\end{document}